\documentclass[%
reprint,
amsmath,amssymb,prx,
aps, longbibliography, superscriptaddress
]{revtex4-1}

\usepackage{graphicx}
\usepackage{dcolumn}
\usepackage{bm}
\usepackage{braket}
\usepackage{mathtools}
\usepackage{bbm}
\usepackage{comment}
\usepackage{makecell}
\usepackage{afterpage}
\usepackage{amsthm}
\usepackage{siunitx}
\usepackage{times}
\usepackage{cellspace}

\usepackage{hyperref}
\hypersetup{
	colorlinks=true,
	linkcolor=blue,
	citecolor=blue,
	filecolor=magenta,
	urlcolor=cyan
}

\def\iu{{\rm i}}

\DeclareMathOperator{\Tr}{Tr}
\def\dif{{\rm d}}
\def\pdif{{\mathcal{D}}}

\usepackage[normalem]{ulem}

\graphicspath{{.}{..}}

\begin{document}
	\title{Anomalous thermal relaxation and pump-probe spectroscopy of 2D topologically ordered systems}
	\date{\today}
	\def\oxf{{Rudolf Peierls Centre for Theoretical Physics, Clarendon Laboratory, Parks Road, Oxford OX1 3PU, United Kingdom}}
	
	\author{Max McGinley}
	\affiliation{\oxf}
	\affiliation{T.C.M. Group, Cavendish Laboratory, JJ Thomson Avenue, Cambridge CB3 0HE, United Kingdom}
	\author{Michele Fava}
	\affiliation{\oxf}
	\affiliation{Philippe Meyer Institute, Physics Department, \'Ecole Normale Sup\'erieure (ENS),
		Universit\'e PSL, 24 rue Lhomond, F-75231 Paris, France}
	\author{S.A. Parameswaran}
	\affiliation{\oxf}
	
	\begin{abstract}
		We study the behaviour of linear and nonlinear spectroscopic quantities in two-dimensional topologically ordered systems, which host anyonic excitations exhibiting fractional statistics. We highlight the role that braiding phases between anyons have on the dynamics of such quasiparticles, which as we show dictates the behaviour of both linear response coefficients at finite temperatures, as well as nonlinear pump-probe response coefficients. These quantities, which act as probes of temporal correlations in the system, are shown to obey distinctive universal forms at sufficiently long timescales. As well as providing an experimentally measurable fingerprint of anyonic statistics, the universal behaviour that we find also demonstrates anomalously fast thermal relaxation: correlation functions decay as a `squished exponential' $C(t) \sim \exp(-[t/\tau]^{3/2})$ at long times. We attribute this unusual asymptotic form to the nonlocal nature of interactions between anyons, which allows relaxation to occur much faster than in systems with quasiparticles interacting via local, non-statistical interactions. While our results apply to any Abelian or non-Abelian topological phase in two-dimensions, we discuss in particular the implications for candidate quantum spin liquid materials, wherein the relevant quantities can be measured using pre-existing time-resolved terahertz-domain spectroscopic techniques.
	\end{abstract}
	
	\maketitle
	
	\section{Introduction}
	
	Strongly correlated many-body systems in two spatial dimensions can host a remarkably rich variety of novel macroscopic quantum phenomena. Perhaps one of the most striking examples is the existence of emergent excitations that exhibit unconventional statistics---so-called `anyons' \cite{Leinaas1977, Wilczek1982}. These quasiparticles are neither bosonic nor fermionic; rather, they possess nontrivial braiding statistics, meaning that the global wavefunction changes when one anyon moves along a path that encircles another. Remarkably, the wavefunction changes in the same way regardless of how far apart the anyons are throughout this process, which points to an effectively nonlocal interaction between excitations. This is only possible in systems whose ground states possess particular patterns of long-ranged entanglement; namely, in 2D topologically ordered phases \cite{Wen1990,Chen2010}.
	
	Over the last several decades, a great deal of progress has been made in understanding the physics of anyons and the topological phases that host them. By now, there are a number of well-known phenomena that are established as being universal to 2D systems possessing excitations with fractional statistics: To name a few, ground state degeneracies appear on surfaces with nonzero genus \cite{Wen1990a}; quantum numbers can fractionalize \cite{Goldstone1981, Laughlin1983}; and the entanglement entropy of large subregions contains a quantized topological contribution \cite{Kitaev2006, Levin2006}. These discoveries each provide important theoretical insight into the nature of topological order, and in some cases also serve as an experimental fingerprint of a given phase of matter.
	
	In addition to the aforementioned properties, which pertain to equilibrium physics, one can also ask about the \textit{dynamics} of systems with anyons. Besides transport measurements (which are challenging in systems with electrically neutral quasiparticles such as quantum spin liquids), the primary means of probing dynamics in solid-state systems is spectroscopy. Theoretical investigations into the behaviour of spectroscopic quantities in topologically ordered systems have begun comparatively recently, and for the most part the focus has been on linear spectroscopy, i.e.~one analyses the signal using linear response theory. For instance, the spin structure factor in quantum spin liquids shows signatures of fractionalization \cite{Cepas2008,Qi2009,Punk2014,Knolle2014a,Knolle2014,Kamfor2014,Knolle2015,Nasu2016}, where excitations must be created in groups of at least two at a time. Similarly, it has been shown how fractional exclusion statistics (a consequence of anyonic statistics, generalizing Pauli's exclusion principle) can imprint themselves in absorption spectra \cite{Morampudi2017}. While these works provide useful insight into the nature of anyon creation and/or annihilation,  there is only so much that can be learned about dynamics from linear response functions, which capture `near-equilibrium' physics. 
	
	In this paper, we reveal universal dynamical phenomena associated with the braiding statistics of quasiparticles in topologically ordered systems, as witnessed by linear and nonlinear spectroscopic quantities. Our primary focus is on pump-probe spectroscopy, where the system is perturbed by a series of two pulses, each of which excite quasiparticles. In particular, we highlight the significance of processes where anyons that were created at different times braid with one another---a possibility that does not arise in linear spectroscopy at zero temperature. As explained in a short paper that serves as a companion to this one \cite{ShortPaper}, such processes dominate the late-time behaviour of the pump-probe response function, and the resulting signal takes a universal form [Eq.~\eqref{eq:main-result}], which constitutes an experimentally measurable signature of anyonic statistics.
	
	One of our aims here is to present concrete calculations that support and generalize the results reported in Ref.~\cite{ShortPaper}, which were justified using more intuitive arguments, most of the time making reference to $\mathbbm{Z}_2$ quantum spin liquids. In brief, by considering the kinematics of those anyons generated by the sequence of pulses, we can compute the probability that their trajectories link in a way that leads to a nonzero braiding phase. Any such process gives a contribution to the pump-probe response coefficient, and this is responsible for the universal form Eq.~\eqref{eq:main-result}. Importantly, since anyons can braid without ever coming close to one another, the probability of braiding is always asymptotically higher than a scattering event due to short-ranged interactions between quasiparticles; therefore, as we shall argue, our result is robust against the inclusion of non-universal local interactions between excitations.
	
	The pump-probe response coefficient is a particularly useful quantity in this context, since it allows one to isolate the effect that a single additional quasiparticles has on the motion of others.  The insight we gain from studying pump-probe spectroscopy is then applied to reveal salient features of linear response functions at finite temperature. Namely, we can consider the probability that anyons created by the time-dependent perturbation braid with thermally activated quasiparticles. Again we find that these processes occur much more often than scattering does, which leads to an anomalously fast decay of the response function in the time domain: a `squished exponential' form is seen $C(t) \sim \exp(-[t/\tau]^{3/2})$, for some temperature-dependent timescale $\tau$ [see Eq.~\eqref{eq:ChiLinFiniteTemp}]. This should be contrasted with the ordinary exponential decay that would be expected from local interactions. Since linear response functions serve as a quantifier of temporal correlations in the system, we conclude that topologically ordered systems exhibit much faster thermal relaxation that systems with quasiparticles having conventional statistics.
	
	While we are not the first to study nonlinear spectroscopy in QSLs and other quantum magnets \cite{Wan2019,Choi2020,Nandkishore2021,Hart2022}, previous works have focused on resolving the homogeneously broadened continuum of fractionalized excitations associated with the creation of multiple excitations, rather than detecting braiding statistics themselves. We also note that fractional exclusion statistics (a consequence of anyonic statistics, generalizing Pauli's exclusion principle) can imprint itself in absorption spectra even in the linear response regime, as discussed in Ref.~\cite{Morampudi2017}. However, the the signal studied in this manuscript probes the braiding of excitations around one another, rather than the physics of their creation. Moreover, given that the pump-probe response coefficient involves the subtraction of two signals, one with a pump pulse and one without [see Eq.~\eqref{eq:ChiDef}], our approach has the advantage that the universal late-time behaviour can be disentangled from non-universal short-distance effects and background contributions, leading to a sharper signal.\\

	Before embarking on any rigorous calculations, we begin our paper by specifying the systems and spectroscopic quantities that are to be studied in this work, and provide intuitive explanation of how the presence of anyonic excitations affects the signal measured in a pump-probe experiment.

	\subsection{Setup and Key Results \label{sec:Setup}}
	
	In this paper, we are concerned with gapped two-dimensional systems, where quasiparticle excitations above the ground state can exhibit generalized statistics: indistinguishable particles can acquire exchange phases that interpolate between fermionic and bosonic, and mutual statistics can even be defined between distinguishable particles. 
	We wish to study the dynamical response of these systems to external probes in regimes beyond linear response, and to understand how the mutual statistics of the lowest energy quasiparticles affects the relevant response coefficients. Primarily, we have in mind both mesoscopic systems in the quantum Hall regime and spin systems that are in (or proximate to) a spin liquid phase.
	
	Our particular focus will be on the response of these systems to pulses of electromagnetic waves. In either of the aforementioned systems, the relevant energy scales correspond to a wavelength of light much greater than any realistic system size. Therefore, we restrict ourselves to external probes that are spatially homogeneous at zero wavevector $k = 0$ -- that is, the operators to which the external electromagnetic fields couple are of the form
	\begin{align}
		\hat{A} = \int \dif^2 \vec{r} \hat{\mathcal{A}}(\vec{r}),
		\label{eq:ZeroMomentum}
	\end{align}
	where $\hat{\mathcal{A}}(\vec{r})$ is a Hermitian operator density. (On a lattice, the integral over space can be replaced by a sum over sites.) Later on, we comment on the possibility of accessing spatially resolved signatures either using inelastic neutron scattering rather than electron spin resonance, or moving to experimental platforms beyond solid state, e.g.~ultracold atoms.
	
	We mainly focus on a particular nonlinear response protocol known as pump-probe spectroscopy. Starting from the ground state (i.e. the quasiparticle vacuum) of the unperturbed Hamiltonian $\hat{H}_0$, $\rho_0 = \ket{\text{VAC}}\bra{\text{VAC}}$, at time $t = 0$ the system is illuminated by a short, intense, `pump' pulse of light which brings the state of the system out of equilibrium. Denoting the operator to which this pulse couples as $\hat{A}_0$, this results in an effectively instantaneous unitary rotation of $\rho_0$
	\begin{align}
		\rho_0 \xrightarrow[\text{pump pulse}]{} e^{-\iu \kappa \hat{A}_0} \rho_0 e^{\iu \kappa \hat{A}_0}
		\label{eq:Pump}
	\end{align}
	for some constant $\kappa$ controlling the strength of the pulse. After a time $t_1$, a second `probe' pulse is applied, whose purpose is to extract properties of the time-evolved non-equilibrium state. Deferring a proper treatment of the probe pulse and the relevant detection schemes to Section \ref{sec:Expt}, for the time being we take it as given that the probe pulse allows one to extract the real part of the dynamical correlator $\braket{\hat{A}_2(t_1+t_2) \hat{A}_1(t_1) }_{\rm pert}$, where the expectation value $\braket{\,\cdot\,}_{\rm pert}$ is taken with respect to the perturbed state in Eq.~\eqref{eq:Pump}, and we work in the interaction picture with respect to $\hat{H}_0$, i.e.~$\hat{A}_1(t_1) = e^{\iu \hat{H}_0 t_1} \hat{A}_1 e^{-\iu \hat{H}_0 t_1}$. The same experiment can be executed without the pump pulse and the results are subtracted to obtain a signal 
	\begin{align}
		\chi_{\rm PP}(t_1, t_2) &= L^{-2}\bigg[\braket{ \hat{A}_2(t_1+t_2)\hat{A}_1(t_1)}_{\rm pert} \hspace*{-2pt} \nonumber\\ &- \braket{ \hat{A}_2(t_1+t_2)\hat{A}_1(t_1)}_0\bigg],
		\label{eq:ChiDef}
	\end{align}
	where the second expectation value is with respect to the original equilibrium state, which is independent of $t_1$, and we have divided by the volume of the system $L^{-2}$ such that $\chi_{\rm PP}$ is intensive. It is common practice in nonlinear spectroscopy to expand the signal in powers of $\kappa$; following standard nomenclature we write $\chi_{\rm PP} = \sum_{n=1}^\infty \kappa^n \chi^{(n+1)}_{\rm PP}$. The coefficients $\chi^{(n+1)}_{\rm PP}$ are nonlinear response functions of second order and higher. In particular, in the present setting the lowest order terms will turn out to be proportional to $\kappa^2$, and therefore we write $\chi_{\rm PP} = \kappa^2 \chi^{(3)}_{PP} + O(\kappa^3)$.
	
	Exact computations of $\chi_{\rm PP}^{(3)}$ that are valid at all times are prohibitively hard, and will depend on the details of the microscopic model in question. Nevertheless, here we argue that in systems where some excitations possess non-trivial braiding statistics, in the long-time limit $t_{1,2} \rightarrow \infty$ the response function $\chi_{\rm PP}^{(3)}$ follows a universal behaviour. In particular, we will show that
	\begin{align}
		\label{eq:main-result}
		\chi_{\rm PP}^{(3)}(t_1, t_2) = c_{\rm PP}  \chi^{(1)}(t_2)\left[ t_2^{3/2} + o(t_2^{3/2})\right]
	\end{align}
	where $\chi^{(1)}(t)=L^{-2}\Tr\left(\hat{A}_2(t) \hat{A}_1(0) \rho_0\right)$ denotes the linear response function, and $c_{\rm PP}$ is a coefficient which depends on the details of the model and can generally be hard to explicitly compute.
	The relationship \eqref{eq:main-result}---which is our main result---is a general feature of 2D systems whose excitations possess non-trivial braiding statistics, and therefore provides a powerful diagnostic tool to characterize fractional statistics using only pulses of light. The magnitude of the subleading term sets a timescale $\tau_{\rm tr}$ beyond which the transient effects represented by the $o(t_2^{3/2})$ term can be safely neglected and the ratio $\chi_{\rm PP}^{(3)}/\chi^{(1)}$ takes its universal form $=c_{\rm PP}t_2^{3/2}$; this timescale will be characterised in later sections, see also Table \ref{tab:Timescales}.

	Most of the manuscript is dedicated to demonstrating the validity of Eq.~\eqref{eq:main-result}, but first we find it instructive to review the following intuitive argument explaining this behaviour, which we reported in Ref.~\cite{ShortPaper}. In the following, and for most of our calculations, we will make explicit reference to systems where all anyons are Abelian, however the non-Abelian case can be treated in much the same way, as we show in Subsection \ref{subsec:NonAb}. \\
	
\begin{figure}
		\centering
		\includegraphics{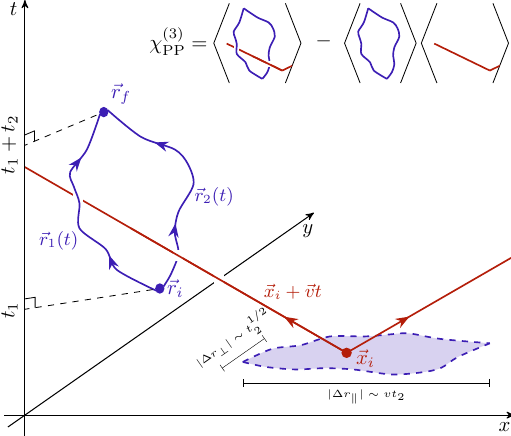}
		\caption{Schematic illustration of the processes contributing to the pump-probe response coefficient \eqref{eq:ChiDef} in a $(2+1)$-dimensional spacetime, using a path integral picture. At time $t = 0$, the pump pulse generates a pair of pump anyons at position $\vec{x}_i$, which in a semiclassical approximation propagate away from one another along trajectories with opposing velocities $\pm \vec{v}$ (red lines). (We omit the backwards-time trajectory in this drawing, which brings these anyons back to their original position $\vec{x}_i$; see Eq.~\eqref{eq:PathIntegralFull}.) A pair of probe anyons is created by the operator $\hat{A}_1$ at time $t_1$, position $\vec{r}_i$, which are later annihilated by $\hat{A}_2$ at time $t_1 + t_2$, position $\vec{r}_f$. In a path integral formalism, the trajectories of the probe anyons are denoted $\vec{r}_{1,2}(t)$, and are drawn as blue lines. Statistical interactions between pump and probe anyons give rise to a phase $e^{2\pi\iu \alpha}$ whenever a pump anyon passes through the loop formed by the probe anyon trajectories. For a fixed $\vec{r}_{1,2}(t)$, we can integrate over all $\vec{x}_i$ such that the paths link. All other contributions cancel upon subtracting the terms in \eqref{eq:ChiDef}, as represented pictorially by the equation in the top right. For trajectories that contribute most to the path integral, the region of $\vec{x}_i$ satisfying this condition (light blue shaded region, dashed outline) has an area that scales as $A \sim t_2^{3/2}$ (see main text). This results in the asymptotic relation \eqref{eq:main-result}, valid in the limit of large $t_{1,2}$.}
		\label{fig:Linking}
	\end{figure}

	Due to their topological nature, quasiparticle excitations with non-trivial mutual statistics can only be created in multiplets of $N > 1$ particles by local operators. Let us focus on $N = 2$ for simplicity, and start by considering the behaviour of the unperturbed two-point function: the second term in Eq.~\eqref{eq:ChiDef}. Since the expectation value is taken with respect to the quasiparticle vacuum, $\hat{A}_1$ must create a quasiparticle pair at time $t_1$ and some position $\vec{r}_i$, which will later be annihilated by $\hat{A}_2$ at time $t_1 + t_2$, position $\vec{r}_f$ [both $\vec{r}_i$ and $\vec{r}_f$ are to be integrated over according to Eq.~\eqref{eq:ZeroMomentum}]. Adopting a path integral formalism for this process, we must integrate over all possible trajectories of these particles $\vec{r}_1(t)$, $\vec{r}_2(t)$ for $t \in [t_1, t_1+t_2]$, weighted by an appropriate action $e^{\iu S[\vec{r}_1(t), \vec{r}_2(t)]}$; these are drawn as blue lines in Fig.~\ref{fig:Linking}. Supposing for now that the quasiparticles are free bosons $S[\vec{r}_a(t)] = (m/2)\int \dif t\,(\dif \vec{r}_a/\dif t)^2$, then the amplitude can be evaluated exactly, and the result is proportional to $e^{-2\iu \Delta t_2}t_2^{-1}$. The frequency of the oscillatory factor $2\Delta$ is the energy required to excite two anyons relative to the quasiparticle vacuum, while the algebraic decay $t_2^{-1}$ reflects the decreasing likelihood of finding two quasiparticles at the same point in space, which is necessary for them to be annihilated.
	
	What changes when the pump pulse is applied beforehand? The post-pump state \eqref{eq:Pump} contains additional quasiparticles, which we refer to as `pump' quasiparticles, to distinguish them from the `probe' excitations created by the probe pulse at time $t_1$. In the absence of interactions (statistical or otherwise), the dynamics of the probe excitations are unchanged by the presence of these pump particles, and so the two terms in \eqref{eq:ChiDef} exactly cancel. Now, suppose that the pump particles have non-trivial braiding statistics with respect to the probe particles. In this case, the action $e^{\iu S[\vec{r}_1(t), \vec{r}_2(t)]}$ must be multiplied by an extra statistical phase, equal to $e^{2\pi \iu \alpha}$ whenever a pump anyon passes through the spacetime loop formed by $\vec{r}_{1,2}(t)$ (see Fig.~\ref{fig:Linking}). Only trajectories that link in this way will contribute to $\chi_{\rm PP}$, since the statistical phase prevents total cancellation of the two terms in \eqref{eq:ChiDef}; this is represented pictorially in the top right inset of Fig.~\ref{fig:Linking}. Therefore, to compute $\chi_{\rm PP}$, we must integrate over $\vec{r}_{1,2}(t)$ as before, but now weighted by the probability that one of the excess pump anyons created by the pump pulse braids with the paths of the probe particles.
	
	Working to leading order in $\kappa$, only a single pair of pump anyons will be created, with the quasiparticles being formed in wavepackets having opposite (crystal) momenta $\vec{k}$, $-\vec{k}$ and being centred around some position $\vec{x}_i$, which again is to be integrated over according to Eq.~\eqref{eq:ZeroMomentum}. These wavepackets propagate away from one another ballistically
	at their group velocities $\pm \vec{v} = \pm \vec{\nabla}_k \epsilon(\vec{k})$, where $\epsilon(\vec{k})$ is the single quasiparticle dispersion \cite{Cyclotron}. The precise distribution of $\vec{k}$ and the dispersion $\epsilon(\vec{k})$ will depend on the microscopic model in question and details of $\hat{A}_0$, but this will not be relevant here; instead we can consider some fixed $\vec{v}$ for now, and leave the averaging over $\vec{v}$ at the end.
	
	Now we must integrate over $\vec{x}_i$. Since the free action is independent of $\vec{x}_i$, this gives a factor equal to the spatial area spanned by those initial positions for which the paths link (blue shaded region in Fig.~\ref{fig:Linking}). The component of $\vec{x}_i$ perpendicular to $\vec{v}$ will be varied over a range of the order of the typical spatial separation of the two probe anyons $\sim |\vec{r}_1(t) - \vec{r}_2(t)|$. By inspecting the free particle action, we see that for typical paths (those for which the phase does not oscillate too rapidly), this distance scales as $\sim \sqrt{t_2/m}$ in the long time limit. In the direction parallel to $\vec{v}$, a shift of $\vec{x}_i$ has the same effect as shifting the spacetime trajectory of the pump anyon upwards in the time direction (see Fig.~\ref{fig:Linking}). Therefore this component should be varied over a range $\sim |\vec{v}|t_2$. Evidently, the space of initial positions $\vec{x}_i$ that yield linking trajectories has an area that asymptotically grows in time as $t_2^{3/2}$. It is this factor, coming from the integral over initial positions of the pump anyon, that leads to the universal form quoted in Eq.~\eqref{eq:main-result}. Note that the average over $\vec{v}$ does not have any bearing on the overall time-dependence; this simply controls the behaviour of the non-universal constant of proportionality $c_{\rm PP}$.
	
	In the particular case we were considering, where $N = 2$ and there are no braiding statistics between the pairs of particles that are created at the same moment in time, we already saw that $\chi^{(1)}(t_2) \propto t_2^{-1}$ up to an oscillatory phase factor, where the decay is due to the decreasing likelihood of anyon recombination. Hence, we have
	\begin{align}
		|\chi_{\rm PP}(t_1, t_2)| \propto \underbrace{\frac{1}{t_2}}_{\text{recombination}} \times \underbrace{t_2^{3/2}}_{\text{spatial integral}} = t_2^{1/2}.
		\label{eq:SqrtT}
	\end{align}
	More generally, if anyons are created in multiplets of $N > 2$ particles, or if there are non-trivial statistics between particles in a given multiplet, then the recombination factor will be modified---see Sections \ref{sec:Calc} and \ref{subsec:MultipletBraiding}. Nevertheless, the $t_2^{3/2}$ factor, which has a purely geometric origin, coming from the integral over $\vec{x}_i$, remains the same. Thus, the relationship \eqref{eq:main-result} is quite general.\\

	While a number of assumptions have been made in this intuitive argument, these are not necessary for the relationship \eqref{eq:main-result} to hold. Most notably, we have so far neglected non-statistical interactions between quasiparticles, and assumed that the system is at exactly zero temperature. In Section \ref{sec:OtherEffects}, we will consider the effects of interactions and finite temperatures more quantitatively, but one can also understand the robustness of our result to such factors at the level of the above argument. Assuming that interactions are sufficiently short-ranged (those decaying faster than $\sim |\vec{r}_1 - \vec{r}_2|^{-\alpha}$ at large separations, with $\alpha > 2$ \cite{Wigner1948}), the presence of pump anyons can only appreciably affect the trajectories of the probe particles when the excitations are closer than some interaction radius $r_{\rm int}$. Using the same geometric approach as before, where one integrates over the initial coordinates of the pump particles keeping the probe anyons' trajectories fixed, the probability of these local scattering processes scales with the perimeter of the loop formed by $\vec{r}_{1,2}(t)$ \cite{Perimeter}. This gives a correction that is subleading compared to the long-ranged statistical interactions, where the relevant probability scales with the area ($t_2^{3/2}$ versus $t_2$).
	
	At finite temperature, the presence of thermally excited quasiparticles (in addition to those created by the pump pulse) modifies the linear response coefficient $\chi^{(1)}(t)$, since braiding between the trajectories of the probe anyons and the thermal excitations leads to an effective dephasing of the two-point correlator $\braket{\hat{A}_2(t)\hat{A}_1(0)}$. However, the pump-probe response function will be modified in exactly the same way. While scattering between thermal and pump quasiparticles may alter the effective distribution of velocities, this only changes $c_{\rm PP}$, and so Eq.~\eqref{eq:main-result} continues to hold. This is shown explicitly later [Eqs.~(\ref{eq:ChiLinFiniteTemp}, \ref{eq:ChiPPFiniteTemp})].

	This concludes our overview of the universal behaviour of the pump-probe response function. In summary, the-late time form of $\chi_{\rm PP}$ obeys a universal relationship Eq.~\eqref{eq:main-result}, which can be understood as described above using a semiclassical picture. The structure of the remainder of our paper is as follows:  To justify our intuitive arguments, in Section \ref{sec:Calc} we compute the main quantity of interest, namely the leading order contribution to $\chi_{\rm PP}(t_1, t_2)$ [Eq.~\eqref{eq:ChiDef}], using an effective low-energy theory for a system with anyonic excitations. In Section \ref{sec:BeyondPert}, we go beyond time-dependent perturbation theory to obtain the full response coefficient at all orders; doing so resolves an apparent paradox that the leading order contribution has an unphysical divergence in the long time limit. In Section \ref{sec:OtherEffects}, we discuss other effects that could not be included in our rigorous calculation, focusing on non-statistical interactions, finite temperatures, and non-Abelian statistics. To make connection between the low-energy theory used before and concrete microscopic models, in Section \ref{sec:Toric} we apply our results to the toric code model in a weak magnetic field, allowing us to connect phenomenological parameters with microscopic quantities. Finally, we discuss how the signal can be measured experimentally in Section \ref{sec:Expt}, before concluding in Section \ref{sec:Concl}.
	
	\section{Calculation of nonlinear response function \label{sec:Calc}}
	
	\subsection{Effective low-energy theory \label{subsec:LowEnergy}}
	
	To begin a calculation of the pump-probe response coefficient, we will require a more detailed characterization of the operators $\hat{A}_{0,1,2}$ appearing in Eqs.~(\ref{eq:Pump}, \ref{eq:ChiDef}), which create and annihilate anyons, as well as a description of how anyons propagate once generated. For the systems we consider in this paper, the lowest-energy excitations are deconfined quasiparticles, which are separated from the ground state by a finite energy gap $\Delta_n > 0$, where the label $n$ is used to distinguish different quasiparticle species. Assuming translation invariance, we can specify a dispersion for each quasiparticle $\epsilon_n(k)$. For the time being, we assume that the only interactions between anyons come through their braiding phases: the wavefunction acquires a phase of $e^{2\pi \iu \alpha_{n n'}}$ when a particle of type $n$ completes a loop that encircles a particle of type $n'$ once in an anticlockwise direction. Later we will include the effect of additional short-range interactions, which do not modify the qualitative form of the response coefficients.
	
	Our analysis applies to 2D topological phases in general, but it will often be helpful to make reference to a particular phase of matter as an example. For this purpose we consider the phase of matter in which the toric code lies \cite{Kitaev1997,Kitaev2003}. Systems in this universality class possess two types of excitations, known as electric and magnetic anyons ($e$ and $m$ respectively). While the electric-electric and magnetic-magnetic braiding phases are trivial $\alpha_{e e} = \alpha_{m m} = 0$, these particles are mutual semions with respect to one another $\alpha_{e m} = \alpha_{m e} = 1/2$. The toric code Hamiltonian is an exactly solvable model with these kind of excitations. At this fine-tuned point, anyons are motionless once created, meaning the dispersion is flat $\epsilon(k) = 0$. However, perturbations that are weak compared to the excitation gap generically induce some dispersion, which endows these excitations with dynamics. In Section \ref{sec:Toric}, we will consider a specific perturbed toric code model, allowing us to relate our universal results to microscopic parameters. 
	
	In general, local operators can only excite quasiparticles in multiplets $\mathcal{N} \coloneqq \{n_1, \ldots,  n_N\}$ that are statistically neutral with respect to all excitations when considered as a composite (i.e.~$\sum_{j=1}^N \alpha_{n_j n'} \in \mathbbm{Z}$ for all $n'$). For example, in the toric code the pairs $\{e, e\}$ and $\{m, m\}$ can be created locally, since braiding two electric anyons around a magnetic anyon gives a trivial phase of $2\pi$. However, individual electric anyons $\{e\}$ cannot be created locally, since they are not neutral with respect to the magnetic anyon. We can associate a threshold energy $\Delta_{\mathcal{N}} = \sum_{j=1}^N \Delta_{n_j}$ to each valid $\mathcal{N}$, which is the minimum energy required to create all the particles in the multiplet. For simplicity we will assume that different multiplets have threshold energies that are well-separated, although we expect that the existence of energetically degenerate multiplets will not wash out the universal signal that we derive here.
	
	The external probes we consider here will have frequencies that are close to these quasiparticle creation thresholds $\Delta_\mathcal{N}$. More formally, writing the microscopic light-matter coupling as a term in the Hamiltonian $f(t) \hat{A}_{\rm micro}$, we take $f(t) = \Re e^{-\iu \omega_0 t} f_0(t)$, where $|\omega_0 - \Delta_\mathcal{N}| \ll \Delta_\mathcal{N}$, and the function $f_0(t)$ varies on a timescale much longer than $\omega_0^{-1}$. While the microscopic operator $\hat{A}_{\rm micro}$ could in principle connect the ground state to complicated states with a larger number of quasiparticles, these components oscillate quickly in the interaction picture, and hence can be ignored (provided one is interested in dynamics on timescales longer than $\Delta_n^{-1}$). After discarding these rapidly oscillating terms, the resulting Hamiltonian only contains operators $\hat{A}_{0,1,2}$ that couple quasiparticle sectors differing by the creation/annihilation of the relevant multiplets. Furthermore, the discrepancy $(\omega_0 - \Delta_\mathcal{N})$ sets an amount of excess kinetic energy that the quasiparticles will have once created. We will assume that this energy is small enough such that the quasiparticle dispersions can be expanded to quadratic order about the band minimum
	\begin{align}
		\epsilon_n(\vec{k}) = \frac{1}{2m_n} \vec{k}^2 + O(k^3)
		\label{eq:QuadDisp}
	\end{align}
	(Anisotropy in the dispersion can also be accounted for in principle, however this will simply result in a rescaling of the pump-probe response function.) We make the above choices in order to progress with our analytical calculation, but we stress that the universal physics discussed here does not depend on the restrictions that we presently impose on the frequency profile of the pump pulse. Indeed, the response to a pulse with a broader range of frequencies will still include the signal we derive here, in addition to non-universal transient effects coming from other mechanisms. In this section, the only interactions between quasiparticles will be due to braiding phases only, and the effect of non-statistical interactions will be treated in Section \ref{subsec:ShortRange}.
	
	Deferring a discussion of the effects of finite temperatures to Section \ref{subsec:FiniteT}, we assume that the system is in its ground state $\rho_0 = \ket{\rm VAC}\bra{\rm VAC}$ before any of the pulses have arrived, i.e.~there are no quasiparticles present. Acting with one of the operators $\hat{A}_{0,1,2}$ on the quasiparticle vacuum, we obtain a perturbed state $\ket{\Psi_{\mathcal{N}}} = \hat{A}_0 \ket{\text{VAC}}$, where $\ket{\Psi_{\mathcal{N}}}$ is some translationally invariant wavefunction in the excitation sector with a single multiplet $\mathcal{N}$. For the time being, we will add one additional restriction, namely that the particles within the multiplets $\mathcal{N}$ created by $\hat{A}_{0,1,2}$ are statistically neutral with respect to one another. This does not preclude nontrivial braiding between excitations in different multiplets $\mathcal{N}$, $\mathcal{N}'$: For example, in the context of the toric code, we can consider $\mathcal{N} = \{e, e\}$ and $\mathcal{N}' = \{m, m\}$. (It will be useful to keep this example in mind in the following.) The reason we make this assumption is that when particles within the set $\mathcal{N}$ possess mutual braiding phases, a short-distance regulator for the operator density $\mathcal{A}(\vec{r})$ appearing in Eq.~\eqref{eq:ZeroMomentum} must be introduced, since such particles cannot be at the same point in space (otherwise the wavefunction would be ill-defined). There is some freedom in choosing this regulator, and non-universal features of the initial $N$-particle wavepacket may affect the subsequent dynamics. We address the more general case in Section \ref{subsec:MultipletBraiding}, but for now we can choose a simple form for $\ket{\Psi_\mathcal{N}}$, where the $N$ pump particles begin in wavepackets localized at the same point in space, i.e. 
	\begin{align}
		\ket{\Psi_\mathcal{N}} \coloneqq \hat{A}_0 \ket{\text{VAC}} \propto \int \dif^2 \vec{x} \ket{\vec{x}} \otimes \cdots \otimes \ket{\vec{x}}
		\label{eq:MultipletCreation}
	\end{align}
	In reality, anyons will not be perfectly pointlike but will have some characteristic size $\xi$ that acts as an ultraviolet cutoff. We will eventually need to invoke this lengthscale to regularize divergent integrals in momentum space, but for now we can assume that anyons generated by each of the perturbing operators $\hat{A}_{0,1,2}$ will be created and annihilated at the same point in space.
	
	The post-pump state takes the form given in Eq.~\eqref{eq:MultipletCreation} for typical zero-momentum operators $\hat{A}_0$ [Eq.~\eqref{eq:ZeroMomentum}]. However, one should bear in mind that in certain scenarios there may be selection rules that further constrain how the system is perturbed by the external pulses, besides those imposed by the fusion rules associated with the underlying topological order. For example, in a spin-half system with unbroken $\mathrm{SU}(2)$ spin-rotation invariance, the only translation invariant operators made up of single-site terms are the total magnetization operators $\hat{M}^\alpha \coloneqq \sum_j\hat{S}_j^\alpha$, where $\hat{S}_j^\alpha$ is the spin operators for lattice site $j$ along the quantization axis $\alpha$; indeed, in electron spin resonance experiments this is the most natural operator to which light will couple. However, since $\hat{M}^\alpha$ generates the symmetry group, the ground state $\ket{\text{VAC}}$ will be unperturbed by the pulse and no signal would be seen \footnote{We thank John Chalker for pointing this out to us.}. In this specific case, one must either account for the small nonzero wavevector of the light pulse, or identify other microscopic operators to which light couples; for instance, the coupling operator describing Raman scattering at $q = 0$ is a spin bilinear, and hence not a symmetry generator \cite{Shastry1990, Cepas2008}. 
	(For smaller symmetry groups, one can always find a polarization of light $\alpha$ such that excited states \eqref{eq:MultipletCreation} are created by $\hat{M}^\alpha$.) From hereon, we will assume that non-symmetry-generating coupling operators $\hat{A}_{0,1,2}$ have been identified, for which the selection rules are not so stringent so as to prevent creation of anyons; thus Eq.~\eqref{eq:MultipletCreation} can be used.\\
	
	Having specified the action of the operators $\hat{A}_{0,1,2}$ within our low-energy effective description, we are now in a position to explicitly calculate response functions, starting with the simplest case, namely linear response.

	\subsection{Warm-up: Linear response \label{subsec:Lin}}
	
	Before embarking on our calculation of the pump-probe response coefficient, it is useful to consider the behaviour of the second term in Eq.~\eqref{eq:ChiDef}, i.e.~the two-time correlation function in the absence of a pump pulse. This is effectively the linear response coefficient $\chi^{(1)}(t) \coloneqq \braket{\hat{A}_2(t)\hat{A}_1(0)}$. A common approach to calculating linear response quantities is to first calculate the Fourier transform of $\chi^{(1)}(t)$ using a spectral representation. 
	In the present setting, the excitations that can be created and annihilated by the operators $\hat{A}_{1,2}$ [which have zero momentum; Eq.~\eqref{eq:ZeroMomentum}] form a $N$-particle continuum spanned by plane wave states $\ket{\vec{k}_1, \ldots, \vec{k}_N}$, subject to the condition $\sum_{j=1}^N\vec{k}_j = \vec{0}$ that is imposed due to conservation of momentum. The spectral density of these states exhibits non-analytic behaviour at a frequency equal to the gap $\Delta_{\mathcal{N}}$. In the simplest case $N = 2$, a stepwise discontinuity appears, and this same kind of discontinuity will generically be present in the Fourier transformed linear response function. Transforming back to the time domain, this behaviour dictates that the late-time form of $\chi^{(1)}(t)$ is proportional to $e^{-\iu \Delta_{\mathcal{N}} t} t^{-1}$, as quoted in Section \ref{sec:Setup}.
	
	Later, we will study the behaviour of the pump-probe response coefficient using a time-domain approach based on semiclassical trajectories. It is therefore worthwhile re-deriving the above form using such a real-time picture. The effect of the operator $\hat{A}_1(0)$ is to create a pair of quasiparticles in the state \eqref{eq:MultipletCreation} at time $t = 0$. The wavefunction of the quasiparticles can be decomposed into wavepackets that have centre of mass position $\vec{x}_i$ and opposing momenta $\vec{k}$ and $-\vec{k}$, where both $\vec{x}_i$ and $\vec{k}$ are to be integrated over \cite{Fava2022}. In the semiclassical limit $\hbar \rightarrow 0$, these quasiparticles propagate away from one another at their group velocity $\vec{v} = \pm \vec{\nabla}_k \epsilon(\vec{k})$, and so their separation grows in time like $2|\vec{v}(\vec{k})|t$.
	
	If we modelled these wavepackets as perfectly pointike (as we would for classical particles), then we would not find any signal for large $t$, since the quasiparticles must be within some fixed distance of each other to be annihilated by the operator $\hat{A}_2(0)$. However, quantum effects lead to a broadening of the profile of these wavepackets: they are not perfectly pointlike, but rather their width grows as $\sim \sqrt{\hbar t/m}$ (restoring $\hbar$ for now). Consequently, at any given time $t$, quasiparticles with momenta that satisfy $2|\vec{v}(\vec{k})|t\lesssim\sqrt{\hbar t/m}$ will have a finite amplitude of annihilation, and so contribute to $\chi^{(1)}(t)$. Expanding $\vec{v}(\vec{k}) \approx \hbar \vec{k}/m$ for small $\vec{k}$, we see that the momenta giving a non-negligible amplitude have a magnitude $\lesssim \sqrt{\hbar m/t}$, and such points occupy an area $\propto 1/t$ in 2D momentum space. If the quasiparticles within this multiplet are mutually bosonic, as in Eq.~\eqref{eq:MultipletCreation}, then the integrand is approximately constant in this region, and we find $\chi^{(1)}(t) \sim 1/t$ as quoted before.
	
	While we will assume trivial statistics within multiplets in the following pump-probe calculation, incidentally we can also use the above picture to understand the behaviour linear response coefficient when there are non-trivial exchange or braiding statistics between quasiparticles created at the same time. In this case we must be more careful in accounting for the matrix elements $\braket{\text{VAC}|\hat{A}_2|\vec{k}, -\vec{k}}\braket{\vec{k}, -\vec{k}|\hat{A}_1|\text{VAC}}$, which controls the distribution of quasiparticle momenta created and annihilated by $\hat{A}_{1,2}$. If the particles are not mutual bosons, Pauli exclusion (or its generalization to anyons) prevents creation of two plane-wave states at the same momentum, and so the matrix element must vanish at $\vec{k} = 0$. For fermions, one readily finds $\braket{\vec{k}, -\vec{k}|\hat{A}_1|\text{VAC}} \sim |\vec{k}|$ at small $|\vec{k}|$, and a calculation analogous to that appearing in Ref.~\cite{Morampudi2017} generalizes this to $|\vec{k}|^{\alpha}$ for anyons (subject to certain conditions on the structure of $\hat{A}_1$; see Section \ref{subsec:MultipletBraiding}). This gives $\chi^{(1)}(t) \propto t^{-1-\alpha}$, which is consistent with the results of Ref.~\cite{Morampudi2017}. Additionally, if $N > 2$ mutually bosonic quasiparticles are created at the same time, then  similar arguments can be used to show $\chi^{(1)}(t) \propto t^{-N + 1}$. The structure of matrix elements for $N > 2$ particles with non-trivial mutual statistics is more complicated; see Ref.~\cite{Morampudi2017}.
	
	Regardless of the intra-multiplet statistics, the key insight to take from the above is that to properly capture the late-time behaviour of the two-time correlator $\braket{\hat{A}_2(t)\hat{A}_1(0)}$, we must account for quantum fluctuations about the semiclassical trajectories, i.e.~the broadening of wavepackets as $\sim \sqrt{\hbar t/m}$.
	
	\subsection{Pump-probe response}
	
	Now we turn to the full pump-probe response coefficient, Eq.~\eqref{eq:ChiDef}, working perturbatively in the strength of the pump pulse $\kappa$. Here, we must distinguish the multiplet $\mathcal{N}$ created by the pump pulse via the operator $\hat{A}_0$ from the multiplet $\mathcal{N}'$ created by the probe pulse operators $\hat{A}_{1,2}$. We assume that the pump and probe pulses have frequency profiles overlapping with the corresponding threshold energies $\Delta_{\mathcal{N}}$ and $ \Delta_{\mathcal{N}'}$, which may be different. Accordingly, we can again infer that each operator either creates or annihilates a multiplet, and so the leading order contributions come at second order in $\kappa$. Following standard naming conventions for nonlinear response coefficients \cite{Mukamel1995}, we define the perturbative response coefficient $\chi^{(3)}_{\rm PP}(t_1, t_2)$ using this expansion
	\begin{align}
		\chi_{\rm PP}(t_1, t_2) &= \kappa^2 \chi^{(3)}_{\rm PP}(t_1, t_2) + O(\kappa^3).
		\label{eq:PPExpandO3}
	\end{align}
	By Taylor expanding the exponentials in Eq.~\eqref{eq:Pump}, we obtain
	\begin{align}
		\chi^{(3)}_{\rm PP}(t_1, t_2) &=
		L^{-2}\bigg(\Tr\left[\hat{A}_2(t_1 + t_2)\hat{A}_1(t_1) \hat{A}_0(0) \rho_0 \hat{A}_0(0)    \right] \nonumber\\ & - \frac{1}{2} \Tr\left[\hat{A}_2(t_1 + t_2) \hat{A}_1(t_1) \{\hat{A}_0(0)^2, \rho_0\}  \right]\bigg).
		\label{eq:PPSecondOrder}
	\end{align}
	These low-order contributions dominate the response in the limit of a weak pump pulse $\kappa \rightarrow 0$, and so we will focus on them for now. However, it is important to bear in mind that the weak pulse limit does not commute with the long time limit $t_2 \rightarrow \infty$, as will be clear once we derive the divergent growth $\chi^{(3)}_{\rm PP} \sim t_2^{1/2}$ [Eq.~\eqref{eq:SqrtT}]. We will remedy this issue in Section \ref{sec:BeyondPert}, where we calculate an expression for $\chi_{\rm PP}(t_1, t_2)$ that includes contributions at all powers of $\kappa$, and thus remains valid as $t_2 \rightarrow \infty$.

	The quantity \eqref{eq:PPSecondOrder} describes a process where a multiplet $\mathcal{N}$ is created at time 0, followed by a multiplet $\mathcal{N}'$ at time $t_1$, which is then annihilated at $t_2$. This is precisely the process that was central to our intuitive argument in Section \ref{sec:Setup} (see Fig.~\ref{fig:Linking}). Using the form of $\hat{A}_0$ given above [Eq.~\eqref{eq:MultipletCreation}], the response coefficient can be written as $\Tr[\zeta\,\hat{A}_2(t_1 + t_2)\hat{A}_1(t_1)]$, where we define
	\begin{align}
		\zeta \coloneqq \ket{\Psi_\mathcal{N}}\bra{\Psi_\mathcal{N}} - \braket{\Psi_\mathcal{N}|\Psi_\mathcal{N}} \ket{\text{VAC}}\bra{\text{VAC}}
		\label{eq:RhoInit}
	\end{align}
	Being unnormalized and not positive-definite, the operator $\zeta$ is not itself a valid density matrix; rather, it includes only the contributions to the pumped state \eqref{eq:Pump} that are second order in $\kappa$. Nevertheless, it is helpful to think of the perturbative response coefficient as the expectation value of $\hat{A}_2(t_1 + t_2)\hat{A}_1(t_1)$ with respect to a `state' $\zeta$, as one would if we were calculating the full response to all orders in $\kappa$. The contributions to this expectation value coming from each of the terms in \eqref{eq:RhoInit} are represented pictorially in the top right inset of Fig.~\ref{fig:Linking}: in the first term, pump and probe anyons are both generated, whereas in the second term, the probe anyons are created on top of the vacuum, and the pump anyons only appear through the multiplicative factor $\braket{\Psi_\mathcal{N}|\Psi_\mathcal{N}}$.
	
	So far, we have not described how the statistical interactions between particles (specifically those between pump and probe excitations) can be included in our description. For this purpose, we find it useful to work in a path integral representation, which we now introduce.

	\subsection{Path integral representation of $\chi_{\rm PP}^{(3)}$ \label{subsec:PathInteg}}
	
	Using Eq.~\eqref{eq:MultipletCreation} and the local form of the operators $\hat{A}_{1,2}$ [Eq.~\eqref{eq:ZeroMomentum}], we can express the response coefficient using a Feynman-Vernon functional integral for the dynamics \cite{Feynman1963}, where both the forward and backward branches of the time evolution in \eqref{eq:PPSecondOrder} are expressed as a sum over paths. We consider all trajectories of the particles in $\mathcal{N}$ between times 0 and $t_1+t_2$, along with those of particles in $\mathcal{N}'$ between times $t_1$ and $t_1 + t_2$. If the only interactions are statistical in nature, then the action can be written as a sum of the free particle actions $S_j[\vec{r}_j(t)] = (m_j/2) \int \dif t\, \dot{r}^2$ plus a topological term $\Lambda[\{\vec{r}(t)\}]$ equal to the cumulative statistical phases associated with the braiding of probe anyons around pump anyons. An explicit formula for $\Lambda[\{\vec{r}(t)\}]$ will not be necessary, however we will later use the fact that $\Lambda$ only depends on the relative coordinates between pump and probe anyons. Overall we have
	\begin{widetext}
		\begin{align}
			\chi_{\rm PP}^{(3)}(t_1, t_2) &= \frac{1}{L^2}\int \dif^2 \vec{x}_{i}^{\,+} \dif^2 \vec{x}_{i}^{\,-} \,\left(\prod_{j=1}^{N'} \dif^2\vec{x}_{f,j}\right)\,\dif^2 \vec{r}_i\,\dif^2 \vec{r}_f  \int_{x_j^\pm(0) = x_i^{\pm}}^{x_j^\pm(t_1+t_2) =x_{f,j }} \left(\prod_{j=1}^N \pdif \vec{x}_j^{\, +}(t) \pdif \vec{x}_j^{\, -}(t) e^{\iu S_j[\vec{x}_j^+] - \iu S_j[\vec{x}_j^-]}\right)
			\nonumber\\ &\times
			\int_{\vec{r}_k(t_1) = \vec{r}_i}^{\vec{r}_k(t_1+t_2)=\vec{r}_f} \prod_{k=1}^{N'} \pdif \vec{r}_k(t) e^{\iu S_k[\vec{r}_k]}
			(e^{\iu \Lambda[\{x_j^+(t) - r_k(t)\}_{j,k}]} - 1)
			\label{eq:PathIntegralFull}
		\end{align}
	\end{widetext}
	Here, $\vec{x}^{\, +}(t)$ and $\vec{x}^{\, -}(t)$ are the trajectories that describe the forward and backwards time evolution in \eqref{eq:PPSecondOrder}, respectively. (For clarity, we consistently use $\vec{x}$ with appropriate subscripts to denote coordinates of pump anyons, and $\vec{r}$ for probe anyons.) Note that no probe anyons are present on the backwards branch, and so the statistical phase $\Lambda$ has no dependence on $\vec{x}_j^{\, -}(t)$ and $\vec{r}_k(t)$. The above expression is an explicit representation of the processes illustrated in Fig.~\ref{fig:Linking} (although the backwards trajectories are not drawn explicitly). The factor of $(e^{\iu \Lambda} - 1)$ arises due to the subtraction of the two terms in $\zeta$ [Eq.~\eqref{eq:RhoInit}]; see the pictorial equation in the top right corner of Fig.~\ref{fig:Linking}.
	
	Unfortunately, exact analytical calculations of the dynamics between times $t_1$ and $t_1 + t_2$ quickly become intractable as the number of particles increases. Even in the minimal case where $|\mathcal{N}| = |\mathcal{N}'| = 2$, the evaluation of the four-body path integral including the statistical interactions does not admit a closed-form solution. However, in the limit of long times $t_{1,2}$, we can make two simplifications. Firstly, at sufficiently large $t_1$ we can consider just one of the pump anyons at a time. We make this approximation on the basis that in the long-time limit, the pump anyons will typically be separated by a large distance, and so the amplitude for the probe anyons braiding around more than one pump anyon is small.
	The result is that the statistical factor $(e^{\iu \Lambda} - 1)$ in \eqref{eq:PathIntegralFull} can be replaced by a sum
	\begin{align}
		(e^{\iu \Lambda} - 1) \rightarrow \sum_{j=1}^N \left(e^{\iu \tilde{\Lambda}_j[\{\vec{x}_j^{\,+}(t) - \vec{r}_k(t)\}_k]} - 1\right)
		\label{eq:TopologicalSum}
	\end{align}
	where the new topological term $\tilde{\Lambda}_j[\{\vec{r}_k(t) - \vec{x}_j(t)\}_k]$ captures the statistical phase associated with the braiding of probe anyons around a single pump anyon $j$  \footnote{The phase $\tilde{\Lambda}_j$ is invariant under gauge transformations by virtue of the fact that all probe anyons are created and annihilated at the same points $\vec{r}_i$ and $\vec{r}_f$, respectively.}.

	Our second simplification is to invoke a stationary phase approximation for the trajectories of the pump anyon $j$. To be specific, we decompose the path $\vec{x}^{+}_j(t)$ into a classical trajectory $\vec{x}_{{\rm cl}, j}(t) = \vec{x}_i^{+} + \vec{v}t$, where $\vec{v}_{j} = (\vec{x}_{f,j} - \vec{x}_i^{+})/(t_1+t_2)$, plus a fluctuating part $\delta \vec{x}^{+}_{j}(t)$, and the free part of the action then becomes $m\vec{v}^2_{j} (t_1+t_2)/2 + S_j[\delta \vec{x}^{+}_{j}]$.
	As we argue in Appendix \ref{app:StationaryPhase}, the dependence of the topological part of the action on $\delta \vec{x}(t)$ can be neglected in the limit of large times, with relative corrections decaying at least as fast as $O(t_{2}^{-1})$, i.e.~we can take the trajectory of the pump anyon to be of constant velocity. The fluctuations $\delta \vec{x}^{+}_{j}(t)$ can then be integrated over, along with the backwards trajectory $\vec{x}^{\,-}_j(t)$ and its initial position $\vec{x}_i^{\, -}$, all of which can be expressed using the Feynman propagator.
	This leaves us with a manageable expression for the pump-probe response function
	\begin{align}
		\chi_{\rm PP}^{(3)}(t_1, t_2) \propto&\; \sum_{j=1}^N \int \dif^2 \vec{v}\,  I_j(\vec{v}, t_2), \text{ where } \label{eq:ChiPPVelInteg}  \\
		I_j(\vec{v}, t_2) \coloneqq& \int \dif^2 \vec{x}_i \,\dif^2 \vec{r}_f \int_{\vec{r}_k(t_1) = 0}^{\vec{r}_k(t_1+t_2)=\vec{r}_f} \prod_{k=1}^{N'} \pdif \vec{r}_k(t)\nonumber\\ \times&  e^{\iu S_k[\vec{r}_k]}\left(e^{\iu \tilde{\Lambda}_j[\{\vec{r}_k(t) - \vec{v}t - \vec{x}_i\}_k]} - 1\right).
		\label{eq:PathIntegApprox}
	\end{align}
	Note that in the regime where the above applies, the pump-probe coefficient has no $t_1$-dependence. This is due to the constant-velocity nature of the pump anyon trajectories, meaning that any change of $t_1 \rightarrow t_1 + \Delta t_1$ can be thought of as equivalent to a rigid shift of $\vec{x}_{{\rm cl}, j}(t) \rightarrow \vec{x}_{{\rm cl}, j}(t) + \vec{v}\Delta t_1$. This is borne out in the above since the path integral over $\delta \vec{x}(t)$ and $\vec{x}^{\,-}_j(t)$ is proportional to $(t_1+t_2)^{-2}$, which cancels with the factor of $(t_1+t_2)^2$ that comes from the change of integration variables from $\vec{x}_f$ to $\vec{v}$. Additionally, the classical contribution to the action $m\vec{v}^2 (t_1+t_2)/2$ cancels with the opposite phase coming from the backwards trajectory, which is why a factor of $e^{\iu m v^2(t_1+t_2)/2}$ does not appear in \eqref{eq:ChiPPVelInteg}.

	Eq.~\eqref{eq:PathIntegApprox} describes the propagator for $N'$ probe particles moving from $\vec{0}$ to $\vec{r}_{f}$ in the presence of a pump anyon whose trajectory is fixed, and given by $\vec{r}_{\rm cl}(t) = \vec{v}t + \vec{x}_i$. Observe that we have made a semiclassical approximation for the path of the pump anyons and not the probe anyons. This is motivated by the insight gained from Section \ref{subsec:Lin}, where we saw that the behaviour of two-time correlation functions requires one to account for fluctuations of the relevant excitations; see also the discussion of Appendix \ref{app:StationaryPhase}.
	
	In Section \ref{subsec:Eval}, we will directly evaluate $I(\vec{v}, t_2)$, but for now it is helpful to briefly make connection with the arguments that we gave in Section \ref{sec:Setup} to justify the scaling form \eqref{eq:SqrtT}. Evidently, the integral over $\vec{x}_i$ in \eqref{eq:PathIntegApprox} is precisely the integral that was responsible for the factor of $t_2^{3/2}$ in \eqref{eq:SqrtT}, and we can move it inside the path integral over $\vec{r}_k(t)$. Being a topological term, $\tilde{\Lambda}$ only takes a finite number of distinct discrete values, and so we can split up the integral $\int \dif^2 \vec{x}_i (e^{\iu \tilde{\Lambda}} - 1)$ into patches where $e^{\iu \tilde{\Lambda}}$ takes different values, to get
	\begin{align}
		\int \dif^2 \vec{x}_i (e^{\iu \tilde{\Lambda}} - 1) = \sum_{c} A_c[\vec{r}_k(t) - \vec{v} t] (e^{\iu \tilde{\Lambda}_c} - 1)
		\label{eq:SumAreas}
	\end{align}
	where $c$ labels the distinct values $\tilde{\Lambda}_c$ that the functional $\tilde{\Lambda}$ can take, and $A_c$ is a functional of $\vec{r}_k(t) - \vec{v} t$, equal to the (unsigned) area in the space of coordinates $\vec{x}_i$ that satisfy $\tilde{\Lambda}[\vec{r}_k - \vec{v} t - \vec{x}_i] = \tilde{\Lambda}_c$.
	
	While we do not have a closed-form expression for $A_c$, the intuitive arguments in Section \ref{sec:Setup} indicate that this should scale as $t_2^{3/2}$, and this will be backed up by our exact calculations. In fact, the scaling of $\chi_{\rm PP}(t_1, t_2)$ can be seen fairly straightforwardly using the geometric interpretation offered by Eq.~\eqref{eq:SumAreas}. First, note that since the only free parameters in this problem are $\vec{v}$, $t_2$, and the quasiparticle masses $m_k$, by dimension counting $I(\vec{v}, t)$ can only depend on velocity and time through the product $|v|\sqrt{t_2}$. Hence, the long-time limit is equivalent to the large-velocity limit. When we take $|\vec{v}| \rightarrow \infty$, the pump anyon will only ever be in the vicinity of the probe anyons for a short $O(v^{-1})$ period of time. The winding number will then be entirely determined by the location of the probe anyons at this instant in time, which we call $\tau$. In the case $N'=2$, the trajectories contributing to the area $A_c$ in Eq.~\eqref{eq:SumAreas} are those where the ray traced by the fast pump anyon passes between two probe anyons at locations $\vec{r}_{1,2}(\tau)$. Thus, the component of $\vec{x}_i$ perpendicular to $\vec{v}$ should be varied over a distance equal to $|{r}_{2, \perp}(\tau) - {r}_{1, \perp}(\tau)|$, where $r_{k,\perp}$ is the component of $\vec{r}_k$ perpendicular to $\vec{v}$, for $k = 1,2$. Varying the component of $\vec{x}_i$ parallel to $\vec{v}$ only changes the collision time $\tau$, and so $I(\vec{v}, t)$ is given by the path integral of $\int_0^{t_2} \dif \tau |{r}_{2, \perp}(\tau) - {r}_{1, \perp}(\tau)|$. By evaluating the integral over trajectories $\vec{r}_{k}(\tau)$, one can show that this quantity is indeed proportional to $t_2^{1/2}$, confirming Eq.~\eqref{eq:SqrtT}.
	
	When $N' > 2$, an similar path integral describing the long-time limit of $I(\vec{v}, t_2)$ can be constructed, but the expression becomes more complicated. To determine $I(\vec{v}, t_2)$ in full generality, and to remove the need to rely on dimension-counting arguments, it is more convenient to return to the Schr{\"o}dinger picture, wherein Eq.~\eqref{eq:PathIntegApprox} can be computed exactly.
	
	\subsection{Evaluating Eq.~\eqref{eq:PathIntegApprox} \label{subsec:Eval}}
	
	Our objective is now to evaluate the function $I_j(\vec{v}, t)$ defined in \eqref{eq:PathIntegApprox}. The action describes $N'$ probe particles propagating in the presence of the pump anyon $j$, which moves along a fixed-velocity trajectory. Since the probe anyons are mutually non-interacting, we can consider the propagator for a single probe anyon $G_k(t_f, \vec{r}_f; t_i, \vec{r}_i) \coloneqq \braket{\vec{r}_f|U_k(t_f;t_i)|\vec{r}_i}$, where $U_k(t_f;t_i)$ is the unitary operator describing time evolution of particle $k$ under the influence of the moving pump anyon from time $t_i$ to $t_f$. In terms of this propagator, we have
	\begin{align}
		I(\vec{v}, t) &= \int \dif^2 \vec{r}_i\,\dif^2 \vec{r}_f \left(\prod_{k=1}^{N'}G_k(t, \vec{r}_f; 0, \vec{r}_i)\right. \nonumber\\ &- \left.\prod_{k=1}^{N'}G_k^{(0)}(t, \vec{r}_f; 0, \vec{r}_i)\right),
		\label{eq:IGreen}
	\end{align}
	where $G_k^{(0)}(t_f, \vec{r}_f; t_i, \vec{r}_i)$ is the propagator without the pump anyon.
	
	Naturally, it is helpful to perform a Galilean boost to a frame moving with velocity $\vec{v}$ relative to the laboratory frame.  We have
	\begin{align}
		G_k(t_f, \vec{r}_f; t_i, \vec{r}_i) = \tilde{G}_k(t_f, \vec{r}_f - vt_f; t_i, \vec{r}_i - vt_i)
		\label{eq:GreenBoost}
	\end{align}
	where $\tilde{G}$ is the propagator in the co-moving frame. In this frame, the pump anyon is static, and so we are free to place it at the origin. We will adopt polar coordinates $(r, \phi)$ with the $x$ axis in the direction of $\vec{v}$.
	
	A standard way to describe the effect of the pump anyon is to introduce an infinitesimally thin flux tube at the origin, whose strength is chosen such that an Aharonov-Bohm phase of $2\pi \alpha_{jk}$ is acquired every time particle $k$ orbits around it. Any vector potential describing such a magnetic field will satisfy $\oint_\Gamma \dif \vec{r} \cdot \vec{A}(\vec{r}) = 2\pi \alpha_{jk}$ for any loop $\Gamma$ circling the origin in an anticlockwise sense. It will be useful to start in the `string gauge', where $\vec{A}(\vec{r})$ is only on the negative $y$-axis, specifically $\vec{A}(\vec{r}) = \Theta(-y)\delta(x) \hat{x}$, where $\hat{x}$ is a unit vector in the direction along $\vec{v}$. We can then perform a gauge transformation $\psi(r, \phi) \rightarrow e^{2\pi\iu\alpha_{jk}\phi }\psi(r, \phi)$, where we restrict $\phi \in (-\pi/2, 3\pi/2]$. This completely eliminates the vector potential at the expense of introducing twisted boundary conditions for all wavefunctions. In particular, wavefunctions can be assumed to be continuous functions of $\phi$ except at $\phi = 3\pi/2$, where we have
	\begin{align}
		\psi(r,-\pi/2 + 0^+) = e^{2\pi\iu  \alpha_j}\psi(r,3\pi/2).
		\label{eq:AnyonicBC}
	\end{align}
	Since the statistical vector potential vanishes in the chosen gauge, the boosted Hamiltonian for particle $k = 1,\ldots, N$ (indexing the probe anyons) becomes
	\begin{align}
		H_{{\rm boost}, k} = \frac{1}{2m_k}\Big[ (p_x - m_kv)^2 + p_y^2\Big] + \frac{1}{2}m_kv^2
		\label{eq:HamBoost}
	\end{align}
	To calculate the propagator for this Hamiltonian, we first have to construct all its eigenstates, subject to the boundary conditions imposed by anyonic statistics \eqref{eq:AnyonicBC}. This is most easily achieved by using a unitary transformation $H'_j = U^\dagger H_{\rm boost} U$, where $U = e^{\iu m_j x v}$ shifts the momentum operator by $m_k v$, which gives $H'_k = \vec{p}^2/2m_k + m_k v^2/2$. In polar coordinates, one obtains $H'_k = p_r^2/2m_k + L^2/2m_kr^2$, where $L = -\iu \partial_\phi$ is the angular momentum operator. The boundary condition \eqref{eq:AnyonicBC} imposes that $L$ must take values of $\ell - \alpha_k$, where $\ell$ is an integer (we drop the label for the pump anyon $j$ on all quantities for the time being). The radial part of the wavefunction must then satisfy Bessel's equation with constant $(\ell - \alpha_k)^2$. The overall solution is
	\begin{align}
		\psi_{q,\ell}(r, \phi) =& \sqrt{\frac{q}{2\pi}}J_{|\ell - \alpha_k|}(qr) e^{\iu(\ell - \alpha_k)\phi} \\ \text{with energy } &E_{q,\ell} = \frac{q^2}{2m_k} + \frac{1}{2}m_kv^2\nonumber
		\label{eq:Eigenstates}
	\end{align}
	which, with the normalization given, form a complete set of states:
	\begin{align}
		\sum_{\ell = -\infty}^\infty \int \dif q\, \psi_{q,\ell}^*(\vec{r}_1) \psi_{q,\ell}(\vec{r}_2) = \delta^{(2)}(\vec{r}_1 - \vec{r}_2)
	\end{align}
	The precise structure of these eigenstates stems from our assumption that the Hamiltonian in the boosted frame is rotationally invariant. This allows us to make analytical progress in the following, but we wish to highlight that the late-time form of the response coefficient will be qualitatively unchanged if rotational symmetry is broken, e.g.~due to anisotropy in the dispersion $\epsilon_n(\vec{k})$. We now have
	\begin{align}
		&\tilde{G}_k(\vec{r}_f, t; \vec{r}_i, 0) = \braket{\vec{r}_f| U e^{-\iu H'_k t} U^\dagger |\vec{r}_i} \nonumber\\
		&= e^{\iu m_k v (x_f-x_i) - \iu m_kv^2t/2}\sum_{\ell = -\infty}^\infty \frac{e^{\iu(\phi_f - \phi_i)(\ell-\alpha_k)}}{2\pi}\nonumber\\ &\times \int_0^\infty q \dif q J_{|\ell-\alpha_k|}(q|r_f|)J_{|\ell-\alpha_k|}(q|r_i|) e^{-\iu q^2 t/2m_k}
	\end{align}
	The integral over $q$ can be evaluated using the standard integral $\int_0^\infty x \dif x\, e^{-p x^2} J_\nu(ax) J_\nu(bx) = (2p)^{-1} e^{-(a^2+b^2)/4p} I_\nu(ab/2p)$ \cite{DLMF}, valid for all $p \in \mathbb{C}$ with $\Re p > 0$, upon setting $p = \iu t/2m_k + 0^+$.
	\begin{align}
		&= \frac{\iu m_k}{2\pi t} e^{\iu m_k v (x_f-x_i) - \iu m_k v^2t/2} \exp\left(-\frac{\iu m_k}{2t + \iu 0^+}(r_f^2 + r_i^2)\right) \nonumber\\ &\times\sum_{\ell = -\infty}^\infty e^{\iu(\ell - \alpha_k)(\phi_f - \phi_i) + \iu \pi|\ell - \alpha_k|/2}J_{|\ell - \alpha_k|}\left(\frac{m_k r_f r_i}{t + \iu 0^+}\right)
		\label{eq:GreenSingleBoosted}
	\end{align}
	The infinitesimal constant ensures that this expression remains valid in the limit $t \rightarrow 0$, and we have used the identity $I_\nu(\iu x + 0^+) = e^{\iu \nu \pi/2}J_\nu(+x)$ for real $x>0$.
	
	Using Eqs.~(\ref{eq:IGreen}, \ref{eq:GreenBoost}), we have
	\begin{align}
		I(v, t_2) &= \int \dif^2 \vec{r}_i \int \dif^2 \vec{r}_f
		\prod_{k = 1}^{N'}\tilde{G}_k\big(t_2,\vec{r}_f -vt_2; t_1,\vec{r}_i\big) \nonumber\\ &- (\text{same with $\alpha_k \rightarrow 0$})
	\end{align}
	which after substituting Eq.~\eqref{eq:GreenSingleBoosted} becomes
	\begin{widetext}
		\begin{align}
			I(v, t_2) &= \frac{-e^{-\iu Mv^2t_2/2}\prod_k (\iu m_k)}{(2\pi t_2)^N}\int_0^\infty r_i \dif r_i \int_0^{2\pi} \dif \phi_i \int_0^\infty  r_f\dif r_f \int_0^{2\pi}\dif \phi_f  e^{\iu M v (r_f\cos\phi_f-r_i\cos\phi_i)}\exp\left(-\frac{\iu M}{2t_2 + \iu 0^+}(r_f^2 + r_i^2)\right) \nonumber\\&\times \prod_{k=1}^{N'} \sum_{\ell_k= -\infty}^\infty e^{\iu(\ell_k - \alpha_k )(\phi_f - \phi_i) + \iu \pi|\ell_k - \alpha_k|/2}J_{|\ell - \alpha_k|}\left(\frac{m_k r_f r_i}{t_2 + \iu 0^+}\right) - (\alpha_k = 0)
			\label{eq:IntegralToEval}
		\end{align}
	\end{widetext}
	where $M = \sum_k m_k$.
	
	Our aim now is to evaluate this integral. As noted previously, on dimensional grounds $I(v, t_2)$ can only depend on velocity and time through the combination $v\sqrt{t_2}$, and so the late-time limit can be understood by considering the behaviour as $v \rightarrow \infty$. More quantitatively, the relevant dimensionless parameter in the problem is $\beta \coloneqq m v^2 t_2$, and so we expect the response to take its asymptotic form when $\beta \gg 1$, i.e.~$t_2 \gg \tau_{\rm tr}$, where $\tau_{\rm tr}$ is a timescale on the order of $1/mv^2$ (see Table \ref{tab:Timescales}). In this limit, the integrand becomes a rapidly oscillating function of $\phi_{i,f}$, which motivates a stationary phase approximation of these integrals. Points of stationary phase occur at $\phi_{i,f} = 0, \pi$, and of the four different combinations, the one that that gives a dominant contribution to $I(v,t_2)$ is $\phi_i = 0$, $\phi_f = \pi$, i.e.~the particles begin far along the positive $x$-axis and drift at a velocity $v$ until they reach the negative $x$-axis. Details of the evaluation of this integral are given in Appendix \ref{app:IntegEval}, the result of which gives
	\begin{align}
		I(v,t_2) \approx e^{\iu(2N-3) \pi/4} \frac{\sqrt{\pi} \prod_k m_k}{16 M^2 (2\pi t_2)^{N'-2}}\Upsilon[\{\alpha_k\}]\sqrt{2M t_2} v  
		\label{eq:IvtFinal}
	\end{align}
	where we have defined a topological quantity
	\begin{align}
		\Upsilon[\{\alpha_k\}] \coloneqq 1 -  (-1)^{\sum_k \alpha_k}\prod_k \cos(\pi \alpha_k).
		\label{eq:UpsilonDef}
	\end{align}
	Performing the necessary integral over $v$ (which should be cut off at large velocity $\sim 1/\xi m$ to account for the finite spread of wavevectors created by the pulse), we obtain a pump-probe response coefficient $\chi_{\rm PP}^{(3)}(t_1, t_2)$ that scales as $t_2^{1/2 - (N'-2)}$. A straightforward calculation gives the linear response coefficient $\chi^{(1)}(t) \propto t_2^{-N'+1}$, and comparing the two we see agreement with the form originally stated in Eq.~\eqref{eq:main-result}.

	\section{Response beyond perturbation theory \label{sec:BeyondPert}}
	
	As we showed in the previous sections, for the case $N' = 2$ the lowest order contributions to the pump-probe response coefficient $\chi_{\rm PP}^{(3)}(t_1, t_2)$ grow as $t_2^{1/2}$ in the limit $t_2 \rightarrow \infty$. The fact that this quantity diverges at late times indicates that a perturbative expansion of the system's response to the external fields begins to fail. Specifically, if the late-time limit is taken while holding $\kappa > 0$ fixed, then higher order terms in Eq.~\eqref{eq:PPSecondOrder} cannot be ignored, and the whole series must instead be resummed. In this section, we derive an expression for the full response of the system without relying on perturbation theory, using arguments that generalize those given above. The result, Eq.~\eqref{eq:ChiPPNonpert}, remains valid in the long time limit for fixed $\kappa$.
	
	When considering higher order contributions to $\chi_{\rm PP}(t_1, t_2)$ [Eq.~\eqref{eq:ChiDef}], the main difference in our analysis is that we must consider the possibility that the pump pulse creates more than a single quasiparticle multiplet. Since the frequency of the pump pulse is tuned close to the threshold energy $\Delta_{\mathcal{N}}$ (which we assume is not close to any other excitation threshold energy), the terms of order $\kappa^{2n}$ will involve the creation of up to $n$ copies of $\mathcal{N}$. For the time being, we will continue to assume that excitations of the system interact with one another only through their statistical interactions, and that there are no nontrivial statistics among particles within either the pump or probe multiplet (generalizations of this scenario are addressed in Section \ref{sec:OtherEffects}). Therefore, if we work in the path integral formalism as in Section \ref{subsec:PathInteg}, for each trajectory of the probe anyons we can identify contributions where a particular number of pump anyons pass through the loop formed by $\vec{r}_{1,2}(t)$, and an appropriate statistical phase can be assigned to each contribution. Specifically, we can separate out processes where $p$ pump anyons pass through the loop from one side, and $p'$ from the opposite side, which yields a phase of $e^{2\pi \iu \alpha(p-p')}$. Our task now is to determine, for each possible trajectory of the probe anyons $\vec{r}_{1,2}(t)$, the probability that the pump anyons follow paths such that this linking condition is satisfied. We denote this probability $Q_{p,p'}[\vec{r}_{1,2}(t)]$. The full non-perturbative response will then be given by the path integral over the probe anyon trajectories weighted by a factor of $\braket{(e^{\iu \Lambda} - 1)}_{\rm pr} \coloneqq \sum_{p,p' = 0}^\infty e^{2\pi \iu \alpha(p-p')}Q_{p,p'}[\vec{r}_{1,2}(t)] - 1$, where again the subtraction of unity is due to the unperturbed correlator in \eqref{eq:ChiDef}. (The angled brackets $\braket{\,\cdot\,}_{\rm pr}$ indicates that this averaging is being performed over the paths of the probe anyons.)
	
\begin{figure}
		\centering
		\includegraphics{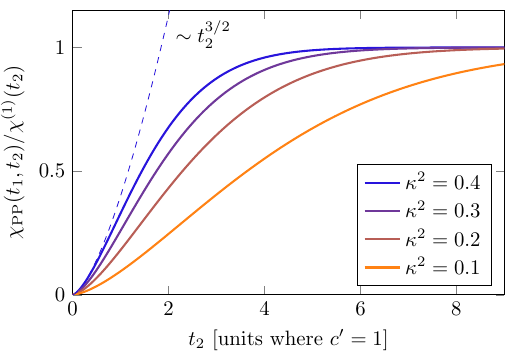}
		\caption{Late-time form of the ratio $\chi_{\rm PP}(t_1, t_2)/\chi^{(1)}(t_2)$, where $\chi_{\rm PP}(t_1, t_2)$ is the full nonlinear response coefficient, including contributions at all orders in $\kappa$, Eq.~\eqref{eq:ChiPPNonpert}. We use units where the non-universal constant $c_{\rm PP} = 1$, and vary $\kappa^2$ from 0.4 (blue) to 0.1 (orange) in steps of 0.1. Initially, the ratio of the response coefficients grow as $t_2^{3/2}$ (dashed line), in agreement with the perturbative expression derived in previous sections, see Eq.~\eqref{eq:main-result}. After some timescale $\tau_{\rm np} \propto \kappa^{-4/3}$, nonperturbative effects become important, and we see a plateau of the ratio.}
		\label{fig:Nonperturb}
	\end{figure}

	Our previous perturbative calculation informs us that the probability to generate a single pump anyon that links with the loop in a particular sense is proportional to $\kappa^2A$, where $A$ is the area functional $A[\vec{r}(t) - \vec{v}t]$ integrated over velocities $\vec{v}$, which comes from integrating over the initial positions of the pump multiplet [see Eq.~\eqref{eq:SumAreas}]. The pump pulse can produce many pump multiplets which are created and propagate approximately independently of one another (assuming their density is low enough), and so the probability that $p$ particles link in a given sense will follow a Poisson distribution, with rate $c \kappa^2 A$ for some constant $c$, i.e.
	\begin{align}
		\text{Prob}(p) = e^{-c\kappa^2 A} \frac{(c\kappa^2A)^p}{p!}
	\end{align}
	Applying the same logic to the paths that link in the opposite sense gives us an expression for $Q_{p,p'}[\vec{r}_{1,2}(t)]$. Thus, the trajectory of each probe anyon should be weighted by a factor
	\begin{align}
		\Braket{(e^{\iu \Lambda} - 1)}_{\rm pr} =& \sum_{p,p'=0}^\infty e^{2\pi \iu \alpha(p-p')} e^{-2c\kappa^2A} \frac{(c\kappa^2A)^{p+p'}}{(p!)(p'!)} \nonumber\\ =&  \exp\left(-2c\kappa^2\big[1 - \cos(2\pi\alpha)\big]A\vphantom{\sum}\right)  - 1
	\end{align}
	Recalling that $A$ is a functional of $\vec{r}_{1,2}(t)$, we must now perform the path integral over the trajectories of the probe anyons. Our previous arguments can be reapplied here, which tell us that for typical paths, $A \propto t^{3/2}$. The full response coefficient is now given by the same path integral expression as the linear response coefficient $\chi^{(1)}(t_2)$, but with the additional weighting of $\Braket{(e^{\iu \Lambda} - 1)}_{\rm pr}$, giving
	\begin{align}
		\chi_{\rm PP}(t_1, t_2) = \chi^{(1)}(t_2) \left[\exp\left(-c_{\rm PP} \kappa^2 t_2^{3/2}\right) - 1\right]
		\label{eq:ChiPPNonpert}
	\end{align}
	where the prefactor in the exponent is identified as the same constant $c_{\rm PP}$ appearing in Eq.~\eqref{eq:main-result}, to ensure agreement with our perturbative results upon expanding \eqref{eq:ChiPPNonpert} to leading order in $\kappa$. Note that $\chi^{(1)}(t_2)$ is bounded in the long-time limit, and so this nonperturbative expression for the pump-probe response coefficient does not diverge, in contrast to $\chi^{(3)}_{\rm PP}$. Evidently, once short-time transient effects have decayed away, the ratio $\chi_{\rm PP}/\chi^{(1)}$ will depend on time only through a universal function of $\kappa^2 t_2^{3/2}$, after choosing units where $c_{\rm PP} = 1$. The factor inside the square brackets in Eq.~\eqref{eq:ChiPPNonpert} is plotted in Fig.~\ref{fig:Nonperturb} for various values of $\kappa$.
	
	The linear response coefficient itself is most easily evaluated in the case where $N' = 2$, and there are no non-trivial braiding phases between anyons created in the same multiplet (this was the case in the toric code example discussed in previous sections). There, one has $\chi^{(1)}(t) \propto t^{-1}$, and hence the pump-probe response takes the form $\chi_{\rm PP}(t_1, t_2) \propto t_2^{-1}(e^{-c_{\rm PP}\kappa^2 t_2^{3/2}} - 1)$. This signal grows as $\sqrt{t_2}$ for times much less than $\tau_{\rm non-pert} \sim (c_{PP} \kappa^2)^{-2/3}$, after which nonperturbative effects become important. At late times, the pump anyons have such a strong effect that the phase coherence of the two-point function is completely lost, and hence the first term in \eqref{eq:ChiDef} completely decays away. This leaves only the second term, which is the unperturbed correlation function, decaying as $t_2^{-1}$. Interestingly, even though the leading order perturbative response coefficient $\chi^{(3)}(t_1, t_2)$ does not diverge when $N' > 2$, our analysis shows that higher order terms, e.g.~$\chi^{(5, 7, \ldots)}$, will always diverge for times beyond $\tau_{\rm non-pert}$; this can be understood as a consequence of the long-ranged nature of the interactions between anyons.

	To summarise, the picture provided by these arguments is that the population of anyons produced by the pump pulse have the effect of dephasing the trajectories of the probe anyons through their mutual statistical interactions. This induces a relative suppression of the two-time correlator compared to its unperturbed value, which leads to a non-zero response coefficient \eqref{eq:ChiDef}. This interpretation will prove useful when we discuss the effects of thermally excited quasiparticles in Section \ref{subsec:FiniteT}.
	
	\section{Robustness to other effects \label{sec:OtherEffects}}

	In our calculation, we have made certain simplifications that allowed us to directly compute the pump-probe response coefficient. Here we consider processes and effects that were neglected above, and demonstrate that the qualitative form of the ratio $\chi_{\rm PP}/\chi^{(1)}$ remains universal in the long-time limit. Specifically, we will discuss the effects of short-ranged interactions (Sec.~\ref{subsec:ShortRange}), finite temperature (Sec.~\ref{subsec:FiniteT}), and non-trivial braiding statistics between within the multiplets that are created by each pulse (\ref{subsec:MultipletBraiding}). We also describe how our analysis can be generalised to systems with non-Abelian anyons in Section \ref{subsec:NonAb}. We will find that a number of timescales emerge from our analysis, which we summarise in Table \ref{tab:Timescales}.
	
	\begin{table}
		\centering
		\begin{ruledtabular}
			\bgroup
			\def\arraystretch{1.2}
			\begin{tabular}{l@{\hspace{0.15in}}cc}
				Timescale & Scaling & Reference \\ \colrule
				Transient effects $\tau_{\rm tr}$ & $1/ m (v^*)^2$ & Section \ref{sec:Calc}\\
				Non-perturbative $\tau_{\rm non-pert}$ & $(c_{\rm PP} \kappa^2)^{-2/3}$ & Eq.~\eqref{eq:ChiPPNonpert} \\
				Pump scattering $\tau_{\rm scat, p}$ & $(v^* \sigma \kappa^2)^{-1}$ & Eq.~\eqref{eq:ChiScatP} \\
				Thermal braiding $\tau_{\rm th}$ & $e^{2\Delta/3T}$ & Eqs.~(\ref{eq:ChiLinFiniteTemp}, \ref{eq:ChiPPFiniteTemp}) \\
				Thermal scattering $\tau_{\rm scat, th}$ & $(v^* \sigma)^{-1} e^{\Delta/T}$ & Section \ref{subsec:FiniteT}
			\end{tabular}
			\egroup
		\end{ruledtabular}
		\caption{Summary of timescales that are relevant to pump-probe spectroscopy, when non-perturbative effects, short-ranged interactions, and finite temperatures are included. Here, $T$ is the temperature, $\kappa$ is the strength of the pump pulse [Eq.~\eqref{eq:Pump}], $v^*$ is the maximum group velocity of quasiparticles, $\sigma$ is the scattering cross section (having dimensions of length in 2D), and $\Delta$ is the gap to excitations. The universal form of the finite-temperature linear response function, Eq.~\eqref{eq:ChiLinFiniteTemp}, can be seen when $\tau_{\rm scat, th} \gg \tau_{\rm th} \gg \tau_{\rm tr}$, which occurs at sufficiently low temperatures. Because thermal effects influence the pump-probe and linear response coefficients in the same way, their ratio remains unchanged; thus to see the universal form \eqref{eq:ChiPPNonpert}, we require only $\tau_{\rm scat, p} \gg \tau_{\rm tr}$ [see Eq.~\eqref{eq:ChiScatP}], which occurs at sufficiently weak pump magnitude.}
		\label{tab:Timescales}
	\end{table}
	
	\subsection{Short-ranged interactions \label{subsec:ShortRange}}
	
	So far we have assumed that the only interactions between quasiparticles are through statistical braiding phases. However, if non-statistical interactions are present, as is the case generically, then the population of quasiparticles created by the pump pulse can influence the two-point correlator measured by the probe pulse through these interactions, and hence modify the response function \eqref{eq:ChiDef}. We argue that when interactions are sufficiently short-ranged, any such effect will be subleading compared to the contribution that we have identified above.
	
	An intuitive way to see this is to employ the path integral perspective that we have used in the previous sections. The effects of short-ranged interactions are only felt by trajectories where a pump anyon comes within some characteristic radius $r_{\rm int}$ of one of the probe anyons, and scatters off it. As before, we can integrate over the initial position of the pump anyons $\vec{x}_i$, keeping all paths otherwise the same [this integral was responsible for the area functional $A_c$ in Eq.~\eqref{eq:SumAreas}]. The range of $\vec{x}_i$ that result in paths where particles come within a distance $r_{\rm int}$ of one another will scale with the perimeter of the probe anyon trajectories, rather than the area of the loop formed by them. The perimeter scales as $t_2$ (see Footnote \cite{Perimeter}), which grows less quickly than the area $\sim t_2^{3/2}$; hence interactions will only modify the subleading contributions to the response coefficient, represented by the term $o(t_2^{3/2})$ in Eq.~\eqref{eq:main-result}.
	
	The above argument provides a relatively straightforward justification of why the late-time form of the perturbative response function $\chi^{(3)}$ should not be altered by short-ranged interactions, but it is also useful to consider a more quantitative approach that does not rely on a perturbative expansion of $\chi_{\rm PP}$. This is particularly important in light of the results of Section \ref{sec:BeyondPert}, where we saw that non-perturbative effects can become important at late times. Looking at the ideal form Eq.~\eqref{eq:ChiPPNonpert}, derived without non-statistical interactions, we see that the universal relationship will remain unchanged if the effects of local scattering between anyons occur on a timescale much longer than $\tau_{\rm non-pert} \coloneqq (c_{\rm PP} \kappa^2)^{-2/3}$. This scattering timescale is defined by the point at which the probability of a scattering event between a pump and probe anyon is order unity. This can be calculated in terms of a scattering cross-section $\sigma$, which in 2D is a length scale: Using standard scattering theory, we have $\tau_{\rm scat, p} = \sigma v^*\lambda_{\rm pu}$, where $\lambda_{\rm pu}$ is the density of anyons created by the pump pulse (which scales as $\kappa^{-2}$), and $v^*$ is a typical velocity of the pump quasiparticles. Naturally, scattering between pump and probe anyons suppresses the two-time correlation function, and so we expect that the ratio $\chi_{\rm PP}/\chi^{(1)}$ will follow the form
	\begin{align}
		\frac{\chi_{\rm PP}(t_1, t_2)}{\chi^{(1)}(t_2)} = \exp\left(-(t_2/\tau_{\rm non-pert})^{3/2} - (t_2/\tau_{\rm scat, p})\right) - 1
		\label{eq:ChiScatP}
	\end{align}
	This modification to Eq.~\eqref{eq:ChiPPNonpert} makes no observable difference if $\tau_{\rm scat, p} \gg \tau_{\rm non-pert}$, i.e.~the statistical interactions alone fully compromise the phase coherence of the probe anyons before scattering processes have had time to take any effect. In fact, as long as $\tau_{\rm scat, p} \gg \tau_{\rm tr}$, then there will be an appropriate window of time in which the universal behaviour \eqref{eq:ChiPPNonpert} can be seen: after transient effects have washed out, but before scattering effects have become appreciable. Note that this is always the case, independently of the system, if the pump pulse is weak enough, viz. $\kappa$ is small enough. Alternatively, having weak interactions or small correlation lengths helps to satisfy this condition for larger values of $\kappa$.
	
	To understand exactly what kinds of interactions count as sufficiently short-ranged, we can revisit the calculation that we described in Section \ref{sec:Calc}. Rotationally symmetric power-law interactions between pump and probe anyons can be included directly into the boosted Hamiltonian \eqref{eq:HamBoost}, and we suppose that at long distances these will decay as $V(\vec{r}_j - \vec{r}_k) \rightarrow V_0 |\vec{r}_j - \vec{r}_k|^{-\gamma}$ for some exponent $\gamma$. In this case, the angular part of the eigenstates \eqref{eq:Eigenstates} will remain unchanged, but the part of the Hamiltonian describing radial motion is now
	\begin{align}
		H_{\rm rad} = \frac{-1}{2m_k}\frac{\dif^2}{\dif r^2} + \frac{(\ell - \alpha)^2}{2m_kr^2} + V(r).
		\label{eq:HamRadInt}
	\end{align}
	If $\gamma > 2$, then by applying dimensional analysis to the above differential operator, we can identify a crossover radius $r_{\rm int} \sim (V_0/2m)^{1/(\gamma - 2)}$ outside of which eigenstates are only weakly modified by the power-law interactions. (This length is not to be confused with the cross section $\sigma$, which would have to be computed via alternative means, e.g.~through solving the appropriate Lippmann-Schwinger equation \cite{Lippmann1950}.) This radius is small for weak interactions, whereas the divergent contributions to the response coefficient are due to processes occurring at large distances $r \gtrsim vt_2$. Hence, for $\gamma > 2$ these interactions will not qualitatively affect the late-time behaviour of the response function.
	
	We do not directly address longer-ranged interactions $\gamma \leq 2$ here, since in this case the assumption that quasiparticles separated by large distances propagate independently is not necessarily true. Indeed, there is no small length scale that can separate the regimes of small and large separation of quasiparticles, and so the key assumptions that went into our argument would be invalidated. It would be interesting to investigate such scenarios in future work, in particular in the context of the fractional quantum Hall effect, where anyons interact via long-ranged Coulomb forces $\gamma = 1$.
	
	\subsection{Finite temperature \label{subsec:FiniteT}}
	
	Another assumption that has been made so far is that the system is in its ground state before the pump pulse arrives. In practice, with the system at finite temperature $T$, a population of thermally excited quasiparticles will be present, which themselves can affect the response of the system to external fields. In the regime $T \ll \Delta$, which we will focus on, the density of this population will be exponentially small $\sim e^{-\Delta/T}$, and so we can safely model the thermal excitations as a dilute gas of weakly interacting quasiparticles.

	Firstly, let us neglect non-statistical interactions and, as a warm-up, consider the linear response coefficient, i.e.~the two-time correlator $\chi_{\rm lin}(t) = \braket{\hat{A}_2(t) \hat{A}_1(0)}$. Focussing on the toric code for concreteness, as before we choose $\hat{A}_{1,2}$ such that a pair of magnetic anyons are create at time $0$ and annihilated at time $t$. The effect of thermal quasiparticles on $\chi_{\rm lin}(t)$ can then be understood using a picture analogous to that presented in Section \ref{sec:BeyondPert}: For each trajectory of the magnetic anyons $\vec{r}_{1,2}(t)$, we can define a probability distribution for how many electric anyons pass through the loop (since $e$ and $m$ and mutual semions, we do not need to distinguish different linking orientations). The difference here is that the electric anyons are thermally activated, instead of being created out of the vacuum by the pump pulse, as before.
	
	Thanks to the diluteness of the quasiparticle gas (the density $\lambda_{\rm th} = \int \dif^2 k/(2\pi)^2 e^{-\epsilon(k)/T}$ scales as $e^{-\Delta/T} \ll 1$), the dynamics of the thermal electric anyons can be safely treated semiclassically \cite{Sachdev1997}. Accordingly, we describe the trajectories of the quasiparticles as straight lines with velocities independently distributed with probability density $P(\vec{v})$, determined by the Boltzmann distribution. Since the electric anyons propagate independently, we can use the same logic as in Section \ref{sec:BeyondPert} to argue that the probability of having $p$ electric anyons linking with the loop formed by $\vec{r}_{1,2}(t)$ follows a Poisson distribution, and in this case the rate is given by $s = \lambda_{\rm th}\int \dif^2 \vec{v}\, P(\vec{v}) A[\vec{r}_{1,2}(t) - \vec{v}t]$, with $A$ the area functional in \eqref{eq:SumAreas}, arising due to the integration over all initial positions of the electric anyons. Thus, each trajectory in the path integral over magnetic anyon trajectories $\vec{r}_{1,2}(t)$ should be weighted by a factor $e^{-s} \sum_p (-1)^p s^p/p! = e^{-2s}$, where $s$ depends on $\vec{r}_{1,2}(t)$ through the area functional.
	
	As usual, for typical trajectories the area functional scales as $t^{3/2}$ at late times, while the density follows an Arrhenius law $\lambda_{\rm th} \sim e^{-\Delta/T}$. Hence, comparing the finite- and zero-temperature response coefficients, we expect to find
	\begin{align}
		\chi^{(1)}(t) = \chi^{(1)}_{T=0}(t)\exp\left(-c\, e^{-\Delta/T} t^{3/2}\right)
		\label{eq:ChiLinFiniteTemp}
	\end{align}
	for some constant $c$. This allows us to define a new timescale $\tau_{\rm th} = (c e^{-\Delta/T})^{-2/3}$ that describes how quickly the braiding phases between thermal and probe anyons degrades two-point functions; see Table \ref{tab:Timescales}. The prefactor may depend on how many anyons are created at a time by the probe pulse, among other factors, but will generally decay algebraically (as $t^{-1}$ for the simple $N'=2$ case considered in previous sections). We see that at small finite temperatures, two-time correlation functions will decay via a characteristic `squished exponential' form $e^{-(t/\tau_{\rm th})^{3/2}}$. Although this unusual form of broadening could in principle serve as a witness of nontrivial braiding even at linear response level, it is likely to be challenging to disentangle from other types of broadening, and as we will see there are constraints on the range of temperatures in which this decay mechanism will be the dominant one. This is why we propose measuring the pump-probe signal, where surplus anyons can be created in a controlled fashion using the pump pulse, and any background signals can be subtracted away according to Eq.~\eqref{eq:ChiDef}.
	
	With the above understood, we can determine the late-time behaviour of the pump-probe response coefficient $\chi_{\rm PP}(t_1, t_2)$ at finite temperature by accounting for both thermal and pump-induced quasiparticles. The perturbed two-point function [the first term in Eq.~\eqref{eq:ChiDef}] is suppressed due to dephasing from both sources of quasiparticles, whereas the unperturbed correlator that is subtracted off has the same form as \eqref{eq:ChiLinFiniteTemp}. The result is
	\begin{align}
		\chi_{\rm PP}(t_1, t_2) &= \chi^{(1)}_{T=0}(t) \exp\left(-c\, e^{-\Delta/T} t^{3/2}_2\right)\times\nonumber\\
		&\times\left[ e^{-c_{\rm PP}\kappa^2 t_2^{3/2}} - 1\right].
		\label{eq:ChiPPFiniteTemp}
	\end{align}
	Comparing \eqref{eq:ChiLinFiniteTemp} and \eqref{eq:ChiPPFiniteTemp}, we see that the universal form of the ratio \eqref{eq:ChiPPNonpert} [which encompasses the perturbative result \eqref{eq:main-result}] continues to hold at finite temperatures, since the linear and pump-probe response coefficients are modulated by the same decaying function. Of course, given the finite sensitivity of detectors in experiments, one wishes to work in a regime where $\tau_{\rm th}$ is large enough such that the individual signals $\chi_{\rm PP}$ and $\chi^{(1)}$ do not become smaller than the experimental resolution before transient effects have worn off. Provided that temperatures can be lowered below $\Delta$, this should be achievable thanks to the exponential dependence of $\tau_{\rm th}$ on $1/T$.
	
	It is interesting to note parallels between these semiclassical arguments and an analogous derivation of the finite-temperature relaxational dynamics of the one-dimensional Ising chain in a transverse field, as studied in Ref.~\cite{Sachdev1997}. In that context, quasiparticles are domain walls of separating domains of opposite magnetization, and so the two-time spin correlator $C(t) = \braket{\hat{Z}_j(t) \hat{Z}_j(0)}$ ($\hat{Z}_j$ is a Pauli spin operator on some site $j$) acquires a phase of $-1$ each time a thermal excitation moves across site $j$. In the dilute-gas regime, when $T$ is much less than the gap to excitations, $C(t)$ is approximately equal its zero zero-temperature value multiplied by a decaying envelope $\sim e^{-t/\tau}$ that accounts for this dephasing due to thermal quasiparticles, which propagate with effectively random trajectories that are governed by the Boltzmann distribution. This multiplicative dephasing factor also arises in our results (\ref{eq:ChiLinFiniteTemp}, \ref{eq:ChiPPFiniteTemp}), with the difference that the mechanism of dephasing is non-local statistical interactions, rather than local scattering phases. This nonlocal mechanism gives rise to an envelope has with a different universal form: $\exp(-(t/\tau_{\rm th})^{3/2})$ instead of an ordinary exponential decay, for some timescale $\tau_{\rm th} \propto e^{2\Delta/3T}$.

	One additional effect that has not yet been accounted for is scattering between the probe anyons and the gas of thermal quasiparticles due to short-range non-statistical interactions. As we saw in the previous section, these scattering processes can lead to a further degradation of the phase coherence of the probe anyons, resulting in additional suppression of the two-time correlators. Assuming that the non-statistical interactions are short-ranged (decaying faster than an inverse square law, as in Section \ref{subsec:ShortRange}), this will result in an ordinary exponential decay $e^{-t_2/\tau_{\rm scat, th}}$, where in analogy to $\tau_{\rm scat, th}$, the characteristic time is given by $\tau_{\rm scat, th} = (v^* \sigma \lambda_{\rm th})^{-1}$, where again $\sigma$ is the scattering cross-section, and $\lambda_{\rm th}$ is the density of thermal quasiparticles. In the dilute gas regime, (low enough temperature and small enough $\kappa$), this envelope should affect the linear and pump-probe response coefficients equally, and hence the ratio $\chi_{\rm PP}/\chi^{(1)}$ should remain unchanged. Moreover, since the ratio $\tau_{\rm scat, th}/\tau_{\rm th}$ grows as $T$ is decreased, at sufficiently low temperatures we will have $\tau_{\rm scat, th} \gg \tau_{\rm th}$, and hence the squished exponential form \eqref{eq:ChiLinFiniteTemp} will also be unaffected.
	
	In addition, the combination of non-statistical interactions and finite temperatures provides a mechanism for the pump anyons to relax towards equilibrium, and this leads to a slow decay of the pump-probe signal with $t_1$. The timescale for this to occur is again very slow due to the diluteness of the thermal excitations, on the order of $\tau_{\rm scat, th}$, and hence it should be possible to find a suitable time delay $t_1$ that is large enough to see the asymptotic form of the response coefficient, but smaller than this thermalization timescale.

	Finally, we remark on the possibility that the quasiparticles themselves may not be stable even at zero temperature, which occurs if the system in question is not actually in a topological phase, but only proximate to one, e.g.~when anyons are weakly confined. In this case, the response coefficient will be altered nontrivially for times $(t_1 + t_2)$ that exceed some cutoff, which is set by either the finite lifetime of quasiparticle excitations (which now remains finite even as $T \rightarrow 0$), or the confinement lengthscale, whichever is reached first. (Note that this affects both the $t_1$ and $t_2$ dependence of $\chi_{\rm PP}$, since the motion of pump anyons is also affected by such effects.) This cutoff diverges as one approaches the transition into the topological phase, and so if the system is proximate enough to a QSL, it will still be possible to observe the universal form described above.
	
	\subsection{Scattering from impurities}
	
	Realistic samples inevitably feature some amount of disorder. This can have two main effects for the dynamics of anyons: (a) impurities or defects can lead to elastic scattering of anyons, and, in certain cases, (b) disorder can generate and trap topological defects (see e.g. Ref.~\onlinecite{PhysRevLett.105.030403}), which have non-trivial braiding properties with the dynamical anyons. In this subsection we discuss the consequences of these two impurity effects on the relaxation of linear and pump-probe response function.
	
	
	\paragraph{Scattering effects.---}
	Although impurities in the sample are static, rather than mobile and dynamic, we can understand the effect of disorder at an approximate level in much the same way as scattering off thermally generated anyons: The impurities realise a short-ranged potential which is felt by the quasiparticles, and can scatter their momenta elastically.  We can define an impurity scattering time $\tau_{\rm imp} = (v^* \sigma \lambda_{\rm imp})^{-1}$, with $\lambda_{\rm imp}$ the density of impurities and $\sigma$ the impurity scattering cross-section. This gives us a typical time scale after which the momentum of a quasiparticle will be appreciably scattered.
	
	Scattering of the probe anyons off impurities will degrade the amplitude for creation and re-annihilation, which will lead to a decay of the pump-probe signal. However, this effect is exactly reproduced in the linear response signal, and hence the ratio $\chi_{\rm PP}/\chi^{(1)}$ will remain unaffected. However, scattering of the pump anyons between times $t_1$ and $t_1 + t_2$ may modify the pump-probe signal in a way that is not counterbalanced by $\chi^{(1)}$. While a detailed calculation of the pump-probe response coefficient in the presence of quenched disorder is beyond the scope of this work, we anticipate that these scattering events will make braiding between pump and probe anyons marginally less likely, since the straight-line trajectories shown in Fig.~\ref{fig:Linking} will have to be modified. The universal signal we describe here will still be observable provided that the timescale $\tau_{\rm imp}$ is longer than the timescale for $t_2$ beyond which transient effects have subsided and the relation \eqref{eq:main-result} becomes valid. Indeed, converting $\tau_{\rm imp}$ to a corresponding mean free path $\ell_{\rm imp}$, we expect such a window of time to exist provided that disorder is not so strong such that $\ell_{\rm imp} \sim a$ where $a$ is the lattice spacing. This is certainly true in any `weak-disorder' regime.
	
	\paragraph{Braiding effects.---}
	The consequences on $\chi^{(1)}(t)$ of defects with nontrivial braiding can be understood along the lines of the argument provided in Subsec.~\ref{subsec:FiniteT} for thermal, i.e. dynamical, anyonic quasiparticles. However, since these topological defects are static, the average number of defects that braids with the anyon pair grows like $(\sqrt{t/m})^2$ --- to be compared with the $vt\times \sqrt{t/m}$ when the thermal excitation have average velocity $v$ --- since $\sqrt{t}$-spreading of the one-particle propagator is now the only contribution to braiding. Consequently, these effects produce a further exponential relaxation of $\chi^{(1)}(t)$ scaling like $\exp(-t/\tau_{\rm an. imp.})$ on top of the faster-than-exponential thermal suppression in Eq.~\eqref{eq:ChiLinFiniteTemp}. Therefore this extra contribution will be subleading for small concentration of impurities.
	
	Instead, regarding the ratio $\chi_{\rm PP}/\chi^{(1)}$, the braiding does not affect the pump anyons: the leading semiclassical contribution is obtained when their trajectories are the same in the forward and backward time evolution, so they cannot braid with the defects. Therefore, following the lines of the arguments in paragraph (a), we see that braiding effects do not impact the ratio $\chi_{\rm PP}/\chi^{(1)}$.

	\subsection{Statistical interactions within multiplets \label{subsec:MultipletBraiding}}
	
	So far, we have considered response functions for perturbations that create multiplets of excitations within which all particles are mutually bosonic. An example that we regularly referred back to was the creation of a pair of electric anyons in the toric code, which have no non-trivial braiding or exchange statistics as a pair, despite being semionic with respect to magnetic excitations. Here we consider what happens if the multiplets created by the pump and/or probe pulses contain excitations that are not bosonic with respect to one another. One example of such a multiplet---again in the context of the toric code---is a pair of electric-magnetic ($em$) composite particles, which are fermionic with respect to one another.
	
	As previously mentioned, an important consequence of non-bosonic statistics within a multiplet is that the constituent excitations cannot exist at the same point in space---a generalization of Pauli's exclusion principle. Thus, we cannot use wavefunctions of the form \eqref{eq:MultipletCreation} as a sensible low-energy description of the state of the system immediately after the pulse. Since the wavefunction must vanish at points where particles coincide, one must invoke a regulator that specifies the limiting behaviour of $\ket{\Psi_{\mathcal{N}}}$ at small separations.
	
	The effect of this generalized exclusion principle can already be seen in linear response functions, as was shown in Ref.~\cite{Morampudi2017}.
	In brief, the authors of that work calculated the dynamical structure factor (the Fourier transform of a two-time correlator $\braket{\text{VAC}|\hat{A}_2(t) \hat{A}_2(0)|\text{VAC}}$) using a low-energy effective theory describing the dynamics of a pair of anyons between times $0$ and $t$. Motivated by lattice models such as the toric code, the regularization of the post-pulse state $\hat{A}_2\ket{\text{VAC}}$ that they chose was a rotationally symmetric wavefunction where the two anyons are separated by a finite exclusion radius $a$, i.e.~$\ket{\Psi_\mathcal{N}} = \int \dif^2 \vec{R} \int_0^{2\pi}\dif \phi \ket{\vec{R}, a, \phi}$, where $\ket{\vec{R}, a, \phi}$ is the two-anyon state with centre of mass $\vec{R}$, and relative displacement $(a,\phi)$ in polar coordinates. Converting their frequency-space results into real time, the late-time behaviour of the correlator follows a power law $t^{-1-\alpha}$, where $\alpha$ is the statistical parameter as before. The same time-dependence can be shown to arise for any uniform state $\ket{\Psi_\mathcal{N}}$ where the initial distance between the two anyons does not exceed some fixed microscopic lengthscale $a$ \footnote{The late-time decay of the linear response function $\braket{\text{VAC}|\hat{A}_2(t) \hat{A}_2(0)|\text{VAC}}$ may in fact be modified if there is a selection rule prohibiting the formation of a two-anyon state with the smallest possible angular momentum. However, since there is no reason to expect such a constraint, we assume that generically the post-pulse state $\hat{A}_2(0)\ket{\text{VAC}}$ has non-zero overlap with this angular momentum sector.}. With the exception of $\alpha = 0$ (bosons), this clearly differs from the $t^{-1}$ linear response behaviour that we argued for in Section \ref{sec:Setup}, which is simply the amplitude for two free particles to recombine [the first factor in Eq.~\eqref{eq:SqrtT}].
	
	While it is clear that individual response functions---linear or otherwise---will be modified by statistical interactions between multiplets, the central quantity in our work is the ratio of the pump-probe and linear response coefficients, which as we argue will continue to follow the universal form derived before Eqs.~(\ref{eq:main-result}, \ref{eq:ChiPPNonpert}). Firstly, the effect of statistical interactions within the pump multiplet will only give rise to a quantitative modification of the distribution of quasiparticle velocities created by the pump pulse: once these quasiparticles are created, they will still propagate ballistically. This only leaves interactions within the probe multiplet. Even with these included, we can still use the path integral representation of the dynamics of probe anyons, described in Sections \ref{sec:Setup}, \ref{sec:Calc}, which tells us that each trajectory of the probe anyons should be weighted by a factor of the area functional $A_c$, equal to the size of the space of initial pump coordinates $\vec{x}_i$ that lead to non-trivial braiding [Eq.~\eqref{eq:SumAreas}]. Crucially even when probe anyons are not mutually bosonic, as was the case considered before, for typical paths this area functional continues to follow the same late-time asymptotic form $A_c \propto t_2^{3/2}$. Accordingly, we still expect Eqs.~(\ref{eq:main-result}, \ref{eq:ChiPPNonpert}) to hold, even though the individual response functions $\chi^{(1)}$, $\chi_{\rm PP}$ are modified. On the basis of the results of Ref.~\cite{Morampudi2017}, in the case where two probe anyons are created at a time $N' = 2$, we expect to see the perturbative pump-probe response coefficient scaling as $\chi_{\rm PP}^{(3)}(t_1, t_2) \propto t_2^{1/2 - \alpha_{\rm pr}}$, where the braiding phase between probe anyons is given by $2\pi\alpha_{\rm pr}$.
	
	The scaling of $A_c$ can be argued for solely using the dimension-counting arguments given at the end of Section \ref{subsec:PathInteg}, where the late-time limit is equated to the limit where the velocity of the pump anyon is taken to be large. At large velocities the area must scale linearly with $v$, and since the only velocity-independent length scale in the problem is $\sqrt{t_2/m}$, and the only dimensionless parameter is $v\sqrt{mt_2}$, this fixes $A_c \propto (t_2/m) \times v\sqrt{mt_2} \propto t_2^{3/2}$. We present a more concrete calculation that confirms this scaling of $A_c$ in Appendix \ref{app:AreaScaling}.

	At the end of this section, we wish to highlight a difference between the results of
	Ref.~\cite{Morampudi2017}, where the effects of particle statistics on linear response coefficients is studied, versus the effect we study in this paper, which shows up only beyond linear response. The former will be seen in systems that possesses fermionic excitations, which have nontrivial exchange statistics, but trivial braiding statistics. In contrast, the universal late time behaviour of the pump-probe response coefficient is a reflection of nontrivial \textit{braiding} statistics: the phase $e^{\iu \Lambda}$ is determined by the linking of paths in spacetime, rather than an exchange of identical particles. Because of this, pump-probe spectroscopy serves as an identifier of topological excitations with braiding statistics, rather than just nontrivial exchange statistics, which arise in non-topological fermionic systems.
	
	\subsection{Non-Abelian statistics \label{subsec:NonAb}}
	
	Until now, we had only made explicit reference to systems with Abelian anyons, where the effect of braiding is to induce a complex phase in the wavefunction. However, our analysis also applies to topological phases whose excitations possess non-Abelian statistics. In such systems, excited states exhibit a topological degeneracy, meaning that an extra discrete quantum degree of freedom is required to fully specify the state of the system, in addition to the positions of the anyons \cite{Nayak2008}. Braiding of excitations results in the application of a unitary rotation acting on this degenerate space. 
	
	These non-Abelian statistical interactions can be incorporated into a path integral language in a similar way to before. In place of the phase $e^{\iu \Lambda}$ in Eq.~\eqref{eq:PathIntegralFull}, we should instead substitute a matrix element of the unitary operator associated with the braid carried out by the trajectories $\vec{x}_j^{\, +}(t)$, $\vec{r}_k(t)$. Specifically, we make the replacement $(e^{\iu \Lambda}-1) \rightarrow \braket{\chi_f|(U[\vec{x}_j^{\,+}(t), \vec{r}_j^{\,+}(t)]] - 1)|\chi_i}$, where $U$ is a functional of the trajectories, depending only on their braiding properties, and $\ket{\chi_{i, f}}$ are discrete wavefunctions in the discrete space, which are set by the specifics of the operators $\hat{A}_{1,2}$ to which the probe pulse couples (see, e.g.~Ref.~\cite{Nayak2008}, Sec. III C).
	
	With the exception of this difference, all our arguments can be applied in exactly the same way as before. In particular, the decomposition of the path integral into topologically distinct contributions [Eq.~\eqref{eq:SumAreas}] still applies, just with the non-Abelian matrix element in place of the complex phase. The functionals $A_c$ depend only on the geometry of the trajectories, and the free part of the action is as before. Thus, the late-time form of the response coefficient should continue to obey the relationship \eqref{eq:main-result}.
	
	\section{Application to perturbed toric code \label{sec:Toric}}
	
	In this section, we study a microscopic Hamiltonian that possesses anyonic excitations, which allows us to apply our general results to a more concrete setup. We are also able to relate the phenomenological parameters used in Section \ref{sec:Calc} (mass $m$, length scale $\xi$, etc.) to properties of the Hamiltonian.
	
	The specific microscopic model that we consider is the toric code perturbed by a magnetic field. In the toric code, qubits are located at the edges $j$ of a square lattice, which we describe using Pauli operators $X_j$, $Y_j$, $Z_j$. The unperturbed Hamiltonian is a sum of four-body terms located at the vertices $v$ and plaquettes $p$ of the lattice \cite{Kitaev1997,Kitaev2003}
	\begin{align}
		\hat{H}_0 = -J_A \sum_{v} \hat{A}_v - J_B \sum_p \hat{B}_p 
	\end{align}
	where the star operators $\hat{A}_v = \prod_{j \in v} \hat{X}_e$ act on all edges around the vertex $v$, and the plaquette operators $\hat{B}_p = \prod_{j \in p} \hat{Z}_p$ act on all edges around a plaquette $p$.
	
	The ground state of $\hat{H}_0$ is the wavefunction stabilized by all star and plaquette operators, $A_v\ket{\text{GS}} = +\ket{\text{GS}}$, $B_p\ket{\text{GS}} = +\ket{\text{GS}}$. Starting from the ground state and acting with $\hat{Z}_e$ on some edge creates a pair of excitations each of energy $J_A$---one for each of the star operators $\hat{A}_v$ that act nontrivially on $e$ and hence anticommute with $\hat{Z}_j$. Similarly, acting with $\hat{X}_j$ creates a pair of excitations on the two plaquettes shared by $j$, each with energy $J_B$. These two types of excitation are referred to as electric ($e$) and magnetic ($m$) anyons respectively. The fact that they are semions with respect to one another can be seen by acting successively with operators $\hat{Z}_j$ in a way that moves the electric particle around a path that encircles a magnetic particle (see Ref.~\cite{Kitaev2003} for details). Since these excited states are exact eigenstates of $\hat{H}_0$, the anyons do not move once created in the absence of any external perturbation. The immobility of the excitations is reflected in the lack of dispersion in the spectrum of $\hat{H}_0$: eigenstates come in highly degenerate multiplets with discrete energies $n_A J_A + n_B J_B$, where $n_A$, $n_B$ are the number of electric and magnetic anyons, respectively.
	
	To endow the anyonic excitations with a dispersion, we introduce a magnetic field, which for simplicity we place in the $x$-$z$ plane. The full Hamiltonian that we consider in this section is
	\begin{align}
		\hat{H} = \hat{H}_0 - h^x \sum_j \hat{X}_j - h^z \sum_j \hat{Z}_j.
	\end{align}
	We work in the limit $J_{A,B} \gg h^{x,z}$. In this limit, we can neglect hybridization of eigenstates with different numbers of magnetic and electric anyons, and the main effect of the magnetic fields is to lift the degeneracy within each excitation number-sector. The $x$-magnetic field generates hopping of magnetic anyons in the dual lattice, and similarly the $z$-magnetic field allows electric anyons to hop in the original lattice. The dispersion of a single electric or magnetic anyon then becomes
	\begin{subequations}
		\begin{align}
			\epsilon^{(e)}(k) &= 2h^z[\cos(k_xa) + \cos(k_ya)] \\
			\epsilon^{(m)}(k) &= 2h^x[\cos(k_xa) + \cos(k_ya)],
		\end{align}
		\label{eq:PTCDispersions}
	\end{subequations}
	\hspace*{-6pt} where $(k_x, k_y)$ is the quasimomentum, and $a$ is the lattice spacing. We have written these dispersions relative to the band minima, which are at energies $\Delta_e = J_A - 4 h^{z}$ and $\Delta_m = J_B - 4 h^x$ (using the same notation for the threshold energies as in Section \ref{subsec:LowEnergy}).

	In a pump-probe experiment, the incoming pulses of light will naturally couple to the microscopic spins. We can choose the polarization of the incoming fields such that the pump pulse couples to the $X$-component of the spins, and the probe pulse couples to the $Z$-component. This way, assuming that the wavelength of the radiation is long compared to the sample size, the time-dependent fields experienced by the system are uniform in space
	\begin{align}
		\hat{V}(t) = B_{\rm pump}(t) \sum_j \hat{X}_j + B_{\rm probe}(t) \sum_j \hat{Z}_j
	\end{align}
	We take the pump pulse to be a Gaussian wavepacket arriving at $t = 0$, centred around a frequency $2\Delta_e + \delta_1$, where $\delta_1$ is a detuning much smaller than $J_B$, with a width of frequencies $1/\tau_1 \ll J_{A,B}$
	\begin{align}
		B_{\rm pump}(t) = \frac{1}{2}B_1 e^{-\iu(2\Delta_e + \delta_1)t - t^2/2\tau_1^2} + \text{c.c.}
	\end{align}
	Due to its frequency profile, the pump pulse can only excite a pair of electric anyons, assuming $J_A$ and $J_B$ are separated by a gap larger than $\tau_1$. We can therefore write down the wavefunction of the system at times $0 < t < t_1$
	\begin{align}
		\ket{\Psi_{e,e}(t)} &= -\iu \int_{-\infty}^t \dif t' B_{\rm pump}(t') 
		\nonumber\\ &\times
		\sum_{j}\sum_{\hat{b} = \hat{x}, \hat{y}} e^{-\iu \hat{H}_{e,e}(t-t')} \ket{j, j+\hat{b}}_{e,e} + O(B_1^2)
	\end{align}
	where $\ket{j, j'}_{e,e}$ is an excited state with electric anyons at lattice sites $j$, $j'$, and $\hat{H}_{e,e}$ is the Hamiltonian in the relevant excitation number-sector. Thanks to the lack of statistical interactions between these two particles, we can compute the time evolution by transforming to plane wave states $\ket{\vec{k}_n}_e = M^{-1}\sum_j e^{\iu \vec{k} \cdot \vec{r}_j}\ket{j}_e $ and using the single-particle dispersion \eqref{eq:PTCDispersions}. Here, $\vec{r}_j$ is the real space coordinate for site $j$, $M$ is the number of sites in the lattice, and the discrete set of wavevectors satisfying periodic boundary conditions are $\vec{k}_n = (2\pi n_x/L, 2\pi n_y/L)$, with $n_{x}, n_y \in \{-M/2+1, \ldots, M/2\}$. We then have
	\begin{align}
		\ket{\Psi_{e,e}(t)} =& \frac{-\iu B_1}{2} \int_{-\infty}^t \dif t'\, e^{-\iu(2\Delta_e+\delta_1)t' - (t')^2/2\tau_1^2}\nonumber\\ \times& M \sum_{n} f(\vec{k}_n) e^{-2\iu [\Delta_e + \epsilon^{(e)}(k_n)](t-t')}\ket{\vec{k}_n, -\vec{k}_n}_{e,e}
		\label{eq:PTCPumpExpand}
	\end{align}
	where the two particle state $\ket{\vec{k}_n, -\vec{k}_n}_{e,e}$ is the wavefunction of a pair of electric anyons in plane wave states with opposing quasimomenta $\vec{k}_n$ and $-\vec{k}_n$, and we have defined $f(\vec{k}) \coloneqq \cos(k_xa) + \cos(k_ya)$. (In performing the time evolution, we have neglected the effective hard-core constraint that two electric anyons cannot reside on the same vertex; however the effect of this is negligible in the regime of interest, as we will see.) The upper limit of the integral over $t'$ can be extended to $+\infty$ for times $t \gg \tau_1$, which gives
	\begin{align}
		\ket{\Psi_{e,e}(t)} &= \frac{-\iu\sqrt{2\pi} B_1 \tau_1}{2} M\sum_n e^{-\tau_1^2(2\epsilon^{(e)}(k_n) - \delta_1)^2/2} \nonumber\\ &\times f(\vec{k}_n) e^{-2\iu t[\Delta_e + \epsilon^{(e)}(k_n)]} \ket{\vec{k}_n, -\vec{k}_n}_{e,e}
		\label{eq:PTCWavPostPump}
	\end{align}
	From this wavefunction we can read off the distribution of quasimomenta of the electric anyons created by the pump pulse. While various different hierarchies of energy scales can in principle be considered, for convenience we will work in a regime where $\delta_1 \ll \tau_1^{-1} \ll h^{z}$, in which case this distribution is peaked near the bottom of the band, allowing us to expand \eqref{eq:PTCDispersions} to quadratic order in $k_{x,y}$. We can therefore consider quadratically dispersing electric anyons with isotropic mass
	\begin{align}
		m_e = \frac{1}{2h^z a^2}.
		\label{eq:PTCMass}
	\end{align}
	The distribution of quasimomenta is then approximately proportional to $e^{-\xi^4_e k^4/2}$, where the length scale is
	\begin{align}
		\xi_e = a\sqrt{2\tau_1 h^z}.
		\label{eq:PTCXi}
	\end{align}
	By transforming back to real space, we find that the wavefunction describes pairs of electric anyons in wavepackets of size $\xi_e$ centred around the same point. This provides a proper UV regularization of the wavefunction \eqref{eq:MultipletCreation} that we employed previously. We observe that $\xi_e$ can be identified as the typical propagation length of the electric anyons over the time window $\tau_1$ during which they are created. Note also that in the regime $\tau_1^{-1} \ll h^z$, we know that $\xi_e$ is much greater than the lattice spacing, which allows us to approximate $f(\vec{k}) \approx 2$. This also justifies our choice to neglect the hard-core constraint on electric anyons in \eqref{eq:PTCPumpExpand}, since components of the wavefunction where two anyons are located at the same vertex are small.
	
	The wavefunction \eqref{eq:PTCWavPostPump} can be used in place of $\ket{\Psi_\mathcal{N}}$ in the operator $\zeta$ defined in Eq.~\eqref{eq:RhoInit}. As in Section \ref{sec:Calc}, we will employ an approximation where we ignore the influence of the magnetic anyons generated by the probe pulse on the trajectories of the original electric anyons. Because of this, when the trace in \eqref{eq:PPSecondOrder} is taken, only contributions where the wavevector on the ket and bra parts of $\zeta$ coincide will survive. We then have
	\begin{align}
		\zeta &= 2\pi \tau_1^2 B_1^2 M^2\sum_n e^{-\xi_e^4 k^4_n}\ket{\vec{k}_n,-\vec{k}_n}\bra{\vec{k}_n, -\vec{k}_n}_{e,e} \nonumber\\
		&+ (\text{terms annihilated by trace}).
		\label{eq:RhoMomentumSum}
	\end{align}

	The probe pulse allows us to measure the two-time correlator appearing in Eq.~\eqref{eq:ChiDef} (see Section \ref{sec:Expt} for details on how this is achieved). In our case, this pulse is polarized along the $x$ axis, which means that the operators $\hat{A}_{1,2}$ are simply $\sum_j \hat{X}_j$. To isolate contributions coming from processes involving two magnetic anyons, the incoming waveform can be frequency-matched to the magnetic anyon pair threshold of $2\Delta_m$, i.e.~the pulse only contains frequency components near this energy. Because of this, we can again expand the magnetic anyon dispersion to quadratic order about the band minimum, and we identify the mass $m_m = (2h^x a^2)^{-1}$. While this assumption is useful for calculations, we expect to see the same qualitative results even if the range of frequencies is broader.
	
	For each term in the sum in \eqref{eq:RhoMomentumSum}, we must compute the two-time correlator $\braket{\hat{A}_2(t_1 + t_2)\hat{A}_1(t_1)}$ of these magnetic anyons in the presence of electric anyons that propagate at the group velocity $v(\vec{k}_n) = \partial_k \epsilon^{(e)}(\vec{k}_n) \approx \vec{v}_n \coloneqq \vec{k}_n/m_e$. The frequency profile of the probe pulse ensures that $\hat{A}_1$ excites a magnetic anyon pair which is de-excited by $\hat{A}_2$. The amplitude for this is precisely the propagator $I(\vec{v}_n, t_2)$ that we computed in Section \ref{subsec:Eval}. Putting everything together, and using the normalization $\braket{\vec{k}_n, -\vec{k}_n|\vec{k}_n, - \vec{k}_n} = M^{-2}$, the long-time limit of the perturbative pump-probe response coefficient becomes
	\begin{align}
		\chi_{\rm PP}^{(3)}(t_1, t_2) &= \frac{2\pi (B_1 \tau_1)^2}{L^2} \sum_n e^{-\xi_e^4 k_n^4} I(\vec{v}_n,t_2) e^{-2\iu \Delta_m t_2} \nonumber\\
		&= 2\pi  (B_1 \tau_1)^2 m_e^2 \nonumber\\ &\times \int \frac{\dif^2 v}{(2\pi)^2} e^{-\xi_e^4 m_e^4 v^4}I(\vec{v},t_2)e^{-2\iu \Delta_m t_2}
	\end{align}
	Using the expression \eqref{eq:IvtFinal}, and restoring the original microscopic quantities using (\ref{eq:PTCMass}, \ref{eq:PTCXi}), we get
	\begin{align}
		\chi_{\rm PP}^{(3)}(t_1, t_2) = \frac{1}{a^2} \frac{\sqrt{\pi}\, \Gamma(3/4)}{256} B_1^2 e^{-2\iu \Delta_m t_2} \sqrt{\frac{\tau_1 t_2}{h^x h^z}}
	\end{align}
	where the factor of $a^{-2}$ arises due to the normalization of $\chi$ by the volume $L^2$, rather than the number of sites $M$. This calculation demonstrates how the universal $t_2^{1/2}$ divergence emerges starting within a specific microscopic model.
	
	To derive this result, we have made certain assumptions about hierarchies of energy scales, namely that the fields $h^{x,z}$ should be weak enough such that hybridization between different anyon sectors is negligible, and that the pulse frequencies are close enough to threshold $\delta_1 \ll h^z$. While deviations from these assumptions may affect the scaling of the prefactor, in general we expect the dependence on $t_2$ to be a universal feature of systems whose excitations possess non-trivial braiding statistics.
	
	\section{Experimental considerations \label{sec:Expt}}
	
	Having studied the behaviour of the pump-probe response coefficient in detail, we now provide a general discussion of the ingredients necessary to measure this quantity in experiment. For most of this section, our focus will be on putative solid-state realizations of quantum spin liquids, for which bulk probes are particularly useful. We comment on other settings later on, namely quantum Hall systems, ultracold atoms and Rydberg atom arrays.
	
	The dynamics of spins in solid state systems typically occur on timescales of order $\sim \SI{1}{\pico\second}$. As an example, in the candidate material $\alpha$-RuCl$_3$, for which there is evidence of a field-induced non-Abelian QSL phase \cite{Yadav2016, Sears2017, Baek2017, Kasahara2018}, the magnetic couplings are estimated to be in the range $70$-$\SI{90}{\kelvin}$ \cite{Banerjee2017}, corresponding to a frequency of $\sim \SI{1.5}{\tera\hertz}$. Recent technical advances have facilitated the generation of high-intensity THz-domain pulses with short time resolution \cite{Blanchard2007,Yeh2007}, which have already been applied to study ultrafast magnetization dynamics in systems with spontaneous macroscopic spin ordering \cite{Yamaguchi2010, Kampfrath2011, Mukai2016, Lu2017}. Here, in analogy with standard pump-probe setups familiar from other kinds of nonlinear spectroscopy \cite{Mukamel1995}, we will describe a sequence of pulses which allows one to measure the particular response coefficient $\chi_{\rm PP}(t_1, t_2)$ [Eq.~\eqref{eq:ChiDef}] in a candidate quantum spin liquid. In fact, this particular sequence has already been used in previous experiments, where the aim was to demonstrate coherent control of spin precessional motion \cite{Yamaguchi2010}. Thus, the effect we describe in this paper should be detectable using currently existing experimental techniques.

	To be specific, we propose to first illuminate the sample with a short intense pump pulse whose frequency range overlaps with the creation threshold energy for a given quasiparticle multiplet (a pair of electric anyons, say). Since the wavelength of THz light is large, the incoming radiation couples directly to the total magnetization $\hat{M}_\alpha$, where the component $\alpha$ is set by the polarization of the magnetic field of the radiation (i.e.~we neglect the momentum of the photons). After waiting for a time $t_1$, a second weaker pulse is applied, which for now we model as infinitely short-lived, giving a magnetic field $B_{\rm pr}(t) = B_0 \delta(t-t_1)$ along a different direction $\beta$. This perturbation modifies the state of the electron spins at later times, and the resulting time-dependent magnetization $\hat{M}_\gamma(t)$ in turn leads to emission of radiation due to free induction decay (FID). The amplitude of the emitted FID radiation can be measured along a chosen polarization $\gamma$ in a time-resolved fashion using e.g.~electro-optic sampling \cite{Nahata1996}, which allows one to infer the time-dependent magnetization $\braket{\hat{M}(t_1 + t_2)}$.
	
	We have already discussed the effect of the pump pulse in Sections \ref{sec:Setup} and \ref{sec:Toric}: the state of the system immediately after the pulse can be described using the right hand side of \eqref{eq:Pump}, where $\hat{A}_0$ includes components of the magnetization operator $\hat{M}$ that oscillate at frequencies within the frequency range of the pulse. As for the probe pulse, since this is weak and infinitesimally short-lived, we can expand to lowest order in $B_0$. If $\rho_{\rm pert}$ is the post-pump state, then immediately after the probe pulse the system is in the state
	$\hat{\rho}_{\rm pert} -\iu B_0[\hat{M}(t_1),  \hat{\rho}_{\rm pert}] + O(B_0^2)$ (we continue to work in the Heisenberg picture, where $\hat{M} = e^{\iu \hat{H} t}\hat{M} e^{-\iu \hat{H} t}$). Then, the expectation value of the magnetization at time $(t_1 + t_2)$ is given by
	\begin{align}
		&\braket{\hat{M}_\gamma(t_1+t_2)}_{B_0} \nonumber\\
		=&\braket{\hat{M}_\gamma(t_1+t_2)}_{B_0 = 0} -\iu B_0 \Tr\Big(\hat{M}_\gamma(t_1 + t_2) \big[\hat{M}_\beta(t_1), \hat{\rho}_{\rm pert}\big] \Big) \nonumber\\
		=& \braket{\hat{M}_\gamma(t_1+t_2)}_{B_0 = 0} + B_0 \Im \Braket{\hat{M}_\gamma(t_1+t_2) \hat{M}_\beta(t_1)}_{\rm pert}
	\end{align}
	Therefore, by extracting the linear-in-$B_0$ part of the magnetization, and subtracting the same quantity without the probe pulse, we obtain the imaginary part of the desired response coefficient \eqref{eq:ChiDef}.
	
	The above procedure is conceptually straightforward and achievable using currently available techniques. Nevertheless, it is also worth contemplating alternative setups that measure the response coefficient directly in the frequency domain, more akin to standard spectroscopic measurements. Rather than using electro-optic sampling to detect the emitted field, one can alternatively perform an absorption measurement, with the detector downstream of the probe pulse, such that the field being detected is a superposition of the probe pulse field and the FID signal $E_{\rm pr}(t) + E_{\rm FID}(t)$. Using a spectrometer, the power spectrum $I(\omega) = |E_{\rm pr}(\omega) + E_{\rm FID}(\omega)|^2$ can be obtained, and since the signal field is weak the signal will be found in the cross-term $2\Re [E_{\rm pr}(\omega)^* E_{\rm FID}(\omega)]$, since the quadratic term $|E_{\rm FID}(\omega)|^2$ can be neglected. This measurement scheme constitutes an intrinsic heterodyne detection of the FID field, with the probe pulse serving as a local oscillator (see Ref.~\cite{Mukamel1995}). Due to the short probe pulse, $E_{\rm pr}(\omega)$ is approximately constant in $\omega$, and so this gives us access to the one-sided Fourier transform of the imaginary part of the response coefficient $\tilde{\chi}_{\rm PP}(t_1, \omega) = \int_0^\infty \dif t_2 e^{\iu \omega t_2}\Im[\chi_{\rm PP}(t_1, t_2)]$, where $t_1$ is set by the time interval between the pump and probe pulses. Since the emitted FID field is $\pi/2$ out of phase with the magnetization \cite{Mukamel1995}, such an experiment would give us direct access to the imaginary part $\Im \tilde{\chi}_{\rm PP}(t_1, \omega)$, and the real part could be reconstructed using the Kramers-Kronig relations.
	
	\begin{table}[]
		\centering
		\begin{ruledtabular}
			\bgroup
			\def\arraystretch{1}
			\begin{tabular}{l c c}
				& Time domain & Frequency domain \\ \colrule & & \\[-4pt]
				\parbox{70pt}{ \raggedright Linear response\\ ($T=0$)} & $t^{-\eta} e^{-\iu \Delta t}$ & $|\delta \omega|^{\eta - 1} \Theta(\delta \omega)$  \\[12pt]
				\parbox{70pt}{ \raggedright Linear response\\ ($T>0$)}  & $\chi^{(1)}_{T=0}(t) e^{-(t/\tau_{\rm th})^{3/2}}$ & $\tilde{\chi}^{(1)}_{T=0}(\omega) \ast \tilde{f}(\omega \tau_{\rm th})$ \\[12pt]
				Pump-probe & $t^{-\eta+3/2}_2 e^{-\iu \Delta t}$  & $|\delta \omega|^{\eta - 5/2} \Theta(\delta \omega)$
			\end{tabular}
			\egroup
		\end{ruledtabular}
		\caption{Summary of the relationships between the linear and pump-probe response coefficients in the time- and frequency-domain; see Eqs.~(\ref{eq:main-result}, \ref{eq:Convolution}). Fourier transforms at frequency $\omega$ are taken with respect to the time $t$ in linear response, and $t_2$ in pump-probe response. We define $\delta \omega \coloneqq \omega - \Delta_{\mathcal{N}'}$ as the frequency relative to the energy threshold for creation of excitations, and the function $\tilde{f}(y)$ is the Fourier transform of $\exp(-|x|^{3/2})$ with respect to $x$, and $\ast$ denotes a convolution. These results are valid in the limit of long times $t$, $t_2$, or sufficiently close to threshold, i.e.~small $\delta \omega$, as appropriate. The exponent $\eta$ depends on the particulars of how anyons are created and annihilated (see Section \ref{subsec:MultipletBraiding}), but the ratio of the linear and pump-probe response coefficients in both the frequency- and time- domain is universal.}
		\label{tab:Frequency}
	\end{table}
	
	Given that there may be scenarios where the measured data is in the frequency domain, let us consider how the universal relationship between linear and pump-probe response coefficients manifests itself in Fourier space. Since Eq.~\eqref{eq:main-result} is valid in the limit of late times, we expect that the relationship will be most stark at frequencies that are close to the non-analytic points of $\tilde{\chi}^{(1)}(\omega)$. In particular, recall from Section \ref{subsec:Lin} that the imaginary part of $\tilde{\chi}^{(1)}(\omega)$---which is proportional to the spectral function of the magnetization operator---exhibits non-analytic behaviour at the threshold frequency $\Delta_{\mathcal{N}'}$, the minimum energy required to create excitations above the quasiparticle vacuum. The nature of the edge singularity in $\tilde{\chi}^{(1)}(\omega)$ will determine the form of non-analytic behaviour seen in $\tilde{\chi}_{\rm PP}^{(3)}(t_1, \omega)$ via the relationship \eqref{eq:main-result}. The simplest case, which applies to all the cases that we have studied in this work, is a power-law singularity, where the linear response coefficient in the time domain follows $\chi^{(1)}(t) \sim \iu t^{-\eta}e^{-\iu \Delta_{\mathcal{N}'}t}$, where the exponent $\eta$ depends on the number of anyons that can be created at a time by the probe pulse, and the statistical phases between them. In frequency space, this gives us
	\begin{align}
		\Im \tilde{\chi}^{(1)}(\omega) \propto \text{sgn}(\omega) \Theta(|\omega| - \Delta_{\mathcal{N}'}) \big| |\omega| - \Delta_{\mathcal{N}'}\big|^{\eta - 1}
	\end{align}
	where the above is expected to hold for $|\omega|$ sufficiently close to $\Delta_{\mathcal{N}'}$. Our time-domain results can be employed to determine the late-time form of the pump-probe response coefficient, which upon Fourier transforming gives
	\begin{align}
		\Im \tilde{\chi}_{\rm PP}^{(3)}(t_1, \omega) \sim \text{sgn}(\omega)\Theta(|\omega| - \Delta_{\mathcal{N}'}) \big| |\omega| - \Delta_{\mathcal{N}'}\big|^{\eta - 5/2}
	\end{align}
	We see a more drastic singularity in the pump-probe response coefficient by virtue of the fact that the ratio $\chi^{(3)}_{\rm PP}(t_1, t_2)/\chi^{(1)}(t_2)$ grows with $t_2$. If $\chi^{(1)}(\omega)$ exhibits more complicated non-analytic behaviour (i.e.~different from a power law), then one can instead use the convolution theorem to determine the corresponding form for $\tilde{\chi}_{\rm PP}^{(3)}(t_1, \omega)$
	\begin{align}
		\tilde{\chi}_{\rm PP}^{(3)}(t_1, \omega) \propto \int_{-\infty}^\infty \frac{\dif \omega'}{2\pi} |\omega'|^{-5/2} \tilde{\chi}^{(1)}(\omega-\omega').
		\label{eq:Convolution}
	\end{align}
	To properly capture the short-time behaviour of $\chi^{(3)}_{\rm PP}$, before the universal signal \eqref{eq:main-result} is dominant, the factor of $|\omega'|^{-5/2}$ should in principle be altered for values of $\omega'$ much larger than $\tau_{\rm tr}^{-1}$. However this will not impact the qualitative form of $\tilde{\chi}_{\rm PP}^{(3)}(t_1, \omega)$ near threshold.
	
	As discussed in Sections \ref{sec:BeyondPert} and \ref{sec:OtherEffects}, at very long times the response functions may be modulated by a decaying envelope due to either non-perturbative effects or suppression due to scattering and/or finite temperatures. In the frequency domain, this results in a `smoothing out' of any non-analytic behaviour over frequency scales on the order $\tau^{-1}$, where $\tau$ is the appropriate timescale (see Table \ref{tab:Timescales}). For example, at finite temperatures the linear and pump-probe response coefficients are modulated by a factor $f(t/\tau_{\rm th})$, where we define $f(x) \coloneqq \exp(-|x|^{3/2})$. Thus, $\tilde{\chi}^{(1)}(\omega)$ will be the convolution of the zero-temperature response function with $\tilde{f}(\omega \tau_{\rm th})$, where $\tilde{f}(y)$ is the Fourier transform of $f(x)$, being a smooth function of $y$ that peaks at $y=0$ and has a width of order unity. In practice, given that such timescales are typically very large (at least for low temperatures and weak pump pulses), it is likely that the measurement apparatus will not be able to resolve these effects, and the formally divergent expressions given above can be used instead. These results are summarised in Table \ref{tab:Frequency}.
	
	We finally remark on some aspects of the generation of the pump anyons. Firstly, in pump-probe spectroscopy, the initial pulse is typically highly intense, with the aim to bring the system strongly out of equilibrium. In this case, assuming that the relevant matrix element for anyon generation [i.e.~the coefficient of proportionality in Eq.~\eqref{eq:MultipletCreation}] is not small, then the density of pump anyons $n_{\rm pump}$ will be fairly high. The pump anyon density can be related to the pulse strength factor $\kappa$ discussed in Section \ref{sec:BeyondPert} as $n_{\rm pump} \propto \kappa^2$, and hence the intensity of the pump pulse controls how long it takes for the nonperturbative regime to set in. We note that our analysis and prediction of the universal form \eqref{eq:ChiPPNonpert} remains valid for finite pump densities, but begins to break down as $n_{\rm pump}$ approaches $1/a^2$ where $a$ is the lattice spacing, i.e.~one pump anyon per unit cell. This regime is unlikely to be reached in practice.
	
	Another possibility is that the key physics described in this work might still be detectable even if the pump anyons are generated incoherently. Indeed, in Section \ref{subsec:FiniteT}, we saw how thermally generated anyons can modify the linear response coefficient. In a sufficiently low temperature regime $e^{-\Delta/T} \ll 1$, such that the scattering time $\tau_{\rm scat}$ is much longer than the dephasing time $\tau_{\rm th}$, we expect to see a characteristic linear response coefficient following Eq.~\eqref{eq:ChiLinFiniteTemp}. Thus, rather than using a pump pulse as a means of generating excess quasiparticles, an increase in temperature could be used. The temperature dependence of frequency-resolved THz absorption measurements at low temperatures could therefore also provide a signature of non-trivial braiding statistics.\\
	
	We conclude this section by addressing other systems that may host topological phases with anyonic excitations. Two-dimensional electron gases in the fractional quantum Hall regime can host Abelian and/or non-Abelian anyons \cite{Stormer1999}; however in practice performing spectroscopy on these systems may be challenging due to the presence of signals coming from other layers of the semiconductor heterostructure devices that are required to realise the electron gas. (In these systems, one can instead use novel device geometries to guide the motion of anyons through edge modes; this approach has recently been employed to detect braiding statistics \cite{Bartolomei2020,Nakamura2020}.) Moreover, since anyons are charged and a magnetic field is present, our analysis would only remain valid up to a timescale set by the cyclotron frequency, see Footnote \cite{Cyclotron}. Outside of the solid state, proposals have been put forward to realise topologically ordered phases in ultracold atomic gases \cite{Paredes2001, Duan2003, Sorensen2005, Cooper2013, Leonard2022}, and more recently in arrays of Rydberg atoms in optical tweezers \cite{Verresen2021}, which have since been implemented in Ref.~\cite{Semeghini2021}. Light-based probes are natural in these settings, and thanks to the high levels of isolation from the environment and the lack of extraneous degrees of freedom, one expects to see clean spectroscopic signatures. Whether the signal we derive here can be seen in this context depends on the system sizes that can be reached, but with large enough samples, nonlinear spectroscopic probes could prove to be a useful probe of anyonic statistics, particularly in platforms where individual atoms cannot be addressed and measured in a spatially resolved way.
	
	\section{Conclusion and outlook \label{sec:Concl}}
	
	We have studied pump-probe spectroscopy of two-dimensional systems that possess excitations with unconventional statistics. Our key result is a universal relationship dictating the late-time behaviour of the response coefficient. The origin of this behaviour can be intuitively understood using a path integral description for the dynamics of quasiparticles: The factor of $t_2^{3/2}$ in Eq.~\eqref{eq:main-result} arises when one calculates the probability that an anyon created by the pump pulse links with the trajectories of anyons created by the probe pulse, see Fig.~\ref{fig:Linking}.
	
	After confirming this result through an explicit calculation of $\chi_{\rm PP}(t_1, t_2)$, we considered the effects of non-statistical short-ranged interactions and finite temperatures, and argued that our result should remain valid even after these effects are included. Accordingly, the relationship between the linear and pump-probe response coefficients \eqref{eq:main-result} serves as a robust fingerprint of anyonic statistics. While our rigorous calculations were performed using a low-energy effective theory for the dynamics of anyons, it is possible to make quantitative connections to specific microscopic models, as we demonstrated for the perturbed toric code. We finally discussed how the relevant signals can be measured using current THz-domain spectroscopic techniques.
	
	Given that the experimental methods necessary to measure the relevant signal are already available, we anticipate that nonlinear spectroscopy could be used to obtain more information about the nature of magnetism in materials that are candidate quantum spin liquids. One of the most actively explored materials in this context is $\alpha$-RuCl$_3$, and neutron scattering and electron spin resonance data provide evidence that under certain applied magnetic fields this system is in or proximate to a QSL phase \cite{Banerjee2016,Wang2017,Ponomaryov2017,Kasahara2018, Wellm2018}. It would therefore be of great interest to investigate the behaviour of the pump-probe response coefficient in microscopic models that are thought to describe the spin dynamics in this material, as well as its close relatives \cite{Takagi2019}. This would allow useful comparison with potential nonlinear spectroscopic experiments on this class of materials. Already, our results indicate that for such a proximate spin liquid, the pump-probe response coefficient should behave in the way discussed above, up to some characteristic timescale dictating the lifetime of quasiparticles, which should diverge close to the transition into a QSL.

	In addition, our work suggests that universal relationships between linear and nonlinear response coefficients may arise in more general topologically ordered systems. For example, in three spatial dimensions excitations can be pointlike or looplike, and mutual statistics between particles and loops can be defined in analogy to the 2D case \cite{Aharonov1959,Alford1989,Krauss1989,Preskill1990}. Understanding how statistical phases between these excitations manifest themselves in nonlinear response will form an interesting direction for future work, which may prove to be useful in the search for topologically ordered materials in higher dimension e.g.~Coulomb spin liquids \cite{Henley2010}.
	
	\section*{Acknowledgements}
	
	We thank Nick Bultinck, Claudio Castelnovo, John Chalker, Rahul Nandkishore, and Steven H. Simon for useful discussions, and Sarang Gopalakrishnan, Romain Vasseur, and Fabian Essler for discussions and collaboration on related work. We acknowledge support from the European Research Council under the European Union Horizon 2020 Research and Innovation Programme, Grant Agreement No. 804213-TMCS, and from the UK Engineering and Physical Sciences Research Council via  grant EP/S020527/1. Statement of compliance with EPSRC policy framework
	on research data: This publication is theoretical work
	that does not require supporting research data
	
	\appendix
	
	\section{Validity of the stationary phase approximation \label{app:StationaryPhase}}
	
	During our calculation of the pump-probe response coefficient in Section \ref{sec:Calc}, we performed a stationary phase approximation for the trajectory of the pump anyon $\vec{r}_j(t)$, which led to the expression \eqref{eq:ChiPPVelInteg}. In this appendix, we provide a concrete justification of this approximation, allowing us to quantify its accuracy.
	
	To begin with, it is helpful to separate out classical paths and fluctuations for the trajectories of all particles, i.e.~we write $\vec{r}_k(t) = \vec{v}\,'t + \vec{r}_i + \delta \vec{r}_k(t)$ for all probe anyons $k$, as well as the trajectories of the pump particles $x_j(t)$ [here, $\vec{v}\,' = (\vec{r}_f - \vec{r}_i)/t_2$]. Using the decomposition of the path integral into topologically distinct sectors, as in Eq.~\eqref{eq:SumAreas}, we have
	\begin{align}
		\chi_{\rm PP}(t_1, t_2) \propto& \int \dif^2 \vec{v}\, \dif^2\vec{v}\,'  e^{\iu v'^2 t_2\sum_k m_k/2} \int \pdif  \big(\delta\vec{x}(t)\big) \nonumber\\ \times&  \int \left(\prod_{k=1}^{N'}\pdif \big(\delta \vec{r}_k(t)\big)  e^{\iu S_0[\delta \vec{r}_k(t)]}\right) e^{\iu S_0[\delta \vec{x}(t)]}  \nonumber\\ \times& \sum_c (e^{\iu \tilde{\Lambda}_c} - 1) A_c[\vec{v}\,''t + \delta \vec{r}_k(t) - \delta \vec{x}(t)]
		\label{eq:ChiAllFluc}
	\end{align}
	where $\vec{v}\,'' = \vec{v}\,' - \vec{v}$, and $A_c$ is a functional of $N'$ trajectories $\vec{r}_k(t)$, equal to the area in the space of coordinates $\vec{x}_i$ that satisfy $\tilde{\Lambda}[\vec{r}_k(t) - \vec{x}_i] = \tilde{\Lambda}_c$ (see the main text). We are interested in the limit of large times $t_{1,2}$, and so it is useful to consider the response coefficient at rescaled times $\chi_{\rm PP}(\lambda t_1, \lambda t_2)$, where $\lambda > 0$ is a dimensionless constant that will be made large. The path integral for this quantity involves trajectories $\delta \vec{x}(s)$ and $\delta \vec{r}_k(s)$, where the new time coordinate $s$ runs over $s \in [0, \lambda(t_1 + t_2)]$ and $[\lambda t_1, \lambda (t_1 +t_2)]$, respectively. For any such trajectory $\delta \vec{x}(s)$, we can define a corresponding trajectory $\delta \vec{x}\,'(t)$ in the original time window $t \in [0,t_1+t_2]$ which takes the form
	\begin{align}
		\delta \vec{x}\,'(t) = \frac{1}{\sqrt{\lambda}} \delta \vec{x}(\lambda t).
		\label{eq:LambdaReparam}
	\end{align}
	A similar transformation for for $\delta \vec{r}_k(s)$ can be made. Crucially, this transformation respects the boundary conditions of the path integral, and leaves the free part of the action $S_0$ invariant, since
	\begin{align}
		\frac{m}{2} \int_0^{t_2} \dif t \, \left(\frac{\dif}{\dif t} \delta \vec{x}\,'(t)\right)^2 = \frac{m}{2} \int_0^{\lambda t_2}  \dif s\, \left(\frac{\dif}{\dif s} \delta \vec{x}(s)\right)^2.
	\end{align}
	Using the reparametrization \eqref{eq:LambdaReparam}, the time-rescaled response coefficient $\chi_{\rm PP}(\lambda t_1, \lambda t_2)$ can be brought into a form identical to the original expression \eqref{eq:ChiAllFluc}, but with the argument of the functional $A_c$ changed to $\vec{v}\,''s + \sqrt{\lambda}[\delta \vec{r}\,'_{\hskip-2pt k}(s/\lambda) - \vec{x}\,'(s/\lambda)]$. In terms of the time coordinate $t \equiv s/\lambda$, this becomes $(\lambda \vec{v}\,'')t + \sqrt{\lambda}[\delta \vec{r}\,'_{\hskip-2pt k}(t) - \vec{x}\,'(t)]$. Now, using the fact that $A_c$ is an area measuring the space of initial coordinates $\vec{x}_i$ that yield a given topological action $\tilde{\Lambda}_c$, we have $A[\kappa \vec{r}_k(t)] = \kappa^2 A[\vec{r}_k]$ for any constant $\kappa > 0$, on geometric grounds. Applying this to the above with $\kappa = 1/\lambda$, we see that the effect of scaling $t_{1,2} \rightarrow \lambda t_{1,2}$ is the same as making the replacement
	\begin{align}
		&A_c[\vec{v}\,''t + \delta \vec{r}_k(t) - \delta \vec{x}(t)] \nonumber\\\rightarrow&\; \text{const.}\times A_c\left[\vec{v}\,''t + \frac{1}{\sqrt{\lambda}}\big(\delta \vec{r}_k(t) - \delta \vec{x}(t)\big)\right]
	\end{align}
	Therefore, expanding $A_c$ as a series in the fluctuations $\delta \vec{x}(t)$, $\vec{r}_k(t)$ becomes an increasingly good approximation as the times $t_{1,2}$ are increased. Specifically, the ratio of the contributions at successive orders is enhanced by a factor of $\lambda^{-1/2}$ under a scaling of time coordinates $t_{1,2} \rightarrow \lambda t_{1,2}$.
	
	If we perform this formal expansion in powers of $\delta \vec{x}(t)$ and $\delta \vec{r}_k(t)$ separately, then all terms that are zeroth order in $\delta \vec{r}_k(t)$ will vanish. This is because such contributions represent processes where all probe anyons move along the same path, and since the probe anyons are statistically neutral as a composite there can be no statistical phase acquired in this case. The leading order term in this expansion will be second order in $\delta \vec{r}_k(t)$ and zeroth order in $\delta \vec{x}(t)$ (since $\delta \vec{r}_k(t) \rightarrow -\vec{r}_k(t)$ is a symmetry of the action). In practice, since we do not have a closed form for $A_c$, it is easier to treat the fluctuations $\delta \vec{r}_k$ exactly, and to set $\delta \vec{x}(t)$ to zero by hand. This is precisely the stationary phase approximation that we made in the main text to obtain the expression \eqref{eq:PathIntegApprox}. Thanks to inversion symmetry, the leading order corrections to this expression will also come at second order in $\delta \vec{x}(t)$, and will hence be $O(t_{1,2}^{-1})$.
	
	Finally, we wish to remark that in deriving the above scaling relation, we have been careful to keep the velocities $\vec{v}$, $\vec{v}\,'$ fixed, even though they are related to real space coordinates $\vec{x}_{i,f}$, $\vec{r}_{i, f}$ in terms of the times $t_{1,2}$ themselves. We keep velocities rather than positions fixed because the upper limits of the velocity integrals will eventually be cut off by a non-universal UV scale $v_{\rm cutoff} \sim 1/\xi m$, where $\xi$ is set by either the lattice constant or the size of the anyon wavepacket, and $v_{\rm cutoff}$ should remain invariant under the scaling transformation.

	\section{Evaluation of Eq.~\eqref{eq:IntegralToEval} \label{app:IntegEval}}
	
	In this appendix we detail how the integral \eqref{eq:IntegralToEval} is evaluated in the limit $v\rightarrow \infty$ by means of a stationary phase approximation. We start by transforming to dimensionless integration variables $u_{i,f} = \sqrt{M/2t_2} r_{i,f}$, and defining the dimensionless parameters $\beta = \sqrt{2t_2M}v$, $\gamma_k = m_k/M$
	\begin{widetext}
		\begin{align}
			&= \left(\frac{\iu}{2\pi t_2}\right)^{N-2}\frac{e^{-\iu M v^2t_2/2}\prod_k m_k }{\pi^2M^2} \int_0^\infty u_i \dif u_i \int_0^{2\pi} \dif \phi_i \int_0^\infty  u_f\dif u_f \int_0^{2\pi}\dif \phi_f  e^{\iu \beta (u_f\cos\phi_f-u_i\cos\phi_i) - \iu(u_i^2+u_f^2)} \nonumber\\&\times \prod_{k=1}^{N'} \sum_{\ell_k= -\infty}^\infty e^{\iu(\ell_k - \alpha_k )(\phi_f - \phi_i) + \iu \pi|\ell_k - \alpha_k|/2}J_{|\ell - \alpha_k|}\left(2\gamma_k u_i u_f\right) - (\alpha_k = 0)
			\label{eq:IDimensionless}
		\end{align}
		As long as $v \neq 0$, the large-$t$ limit of the above can be extracted by taking the limit $\beta \rightarrow \infty$. If $v = 0$ then all $t_2$-dependence drops out, and we obtain a contribution that is constant in $t_2$. This contribution we ignore for now.
		From here on we assume $v \neq 0$, and take the long time limit via $\beta \rightarrow \infty$; this is valid for $t_2 \gg 1/v^2M$. Additionally, since the statistical parameters $\alpha_k$ are only defined modulo an integer, we can without loss of generality choose $\alpha_k \in [0,1)$.\\

		Since $\beta$ is large, the angular integrals can be evaluated using a stationary phase approximation, with stationary points at $\phi_{i,f} = 0, \pi$. This gives
		\begin{align}
			\left(\frac{\iu}{2\pi t_2}\right)^{N-2} \frac{2e^{-\iu Mv^2t_2/2}\prod_k m_k}{\pi \beta M^2} &\int_0^\infty u_i^{1/2} \dif u_i \int_0^\infty u_f^{1/2} \dif u_f e^{-\iu(u_i^2 + u_f^2)} \sum_{\sigma_i,\sigma_f = \pm 1} e^{-\iu\pi (\sigma_f-\sigma_i)/4} e^{\iu \beta(x_f \sigma_f - x_i \sigma_i)} \left[\prod_k A_k^{(\alpha_k)} - \prod_k A_k^{(0)}\right]\label{eq:Lab} \\
			\text{where } A_k^{(\alpha_k)} \coloneqq& e^{-\iu \pi \alpha_k(\sigma_i-\sigma_f)/2} \sum_{\ell=1}^\infty(\sigma_i\sigma_f)^\ell\left[e^{\iu\pi(\ell - \alpha_k)/2}J_{\ell - \alpha_k}(2\gamma_ku_iu_f) + \sigma e^{\iu \pi(\ell - [1-\alpha_k])/2}J_{\ell - [1-\alpha_k]}(2\gamma_ku_iu_f)  \right] \nonumber
		\end{align}
	\end{widetext}
	The sums over $\sigma_{i,f}$ are for the different stationary points at $\phi_{i,f} = 0, \pi$, and we have split up the sums over $\ell$ into separate parts where $\ell_k-\alpha_k$ is either positive or negative.
	Now we evaluate the sums using Ref.~\cite{Prudnikov1986} Eq.~5.7.5.1, which can be manipulated to give
	\begin{align}
		\sum_{l=1}^\infty (\iu \sigma)^l J_{l-\alpha}(z) &= \frac{1}{2}\int_0^z \dif u \Big[\iu \sigma J_\mu(z-u)J_{-\mu -\alpha}(u) \nonumber\\ &- J_\mu(z-u) J_{1-\mu-\alpha}(u) \Big]
		\label{eq:LSum}
	\end{align}
	where $\mu$ can be any real value satisfying $-1 < \mu  < 1-\alpha$.
	
	Now, we note that the integrand in \eqref{eq:Lab} is a fast-oscillating function of $u_f$ and $u_i$, and hence will be dominated by contributions at large $u_f, u_i \gtrsim \beta$, where the Bessel functions oscillate equally quickly. Therefore, we can take the large-$z$ limit of \eqref{eq:LSum}, which simplifies using the asymptotic form $J_\mu(u) \approx \sqrt{2/\pi u}\cos(u - \mu\pi/2 - \pi/4)$, valid for large real positive $u$:
	\begin{widetext}
		\begin{align}
			&\frac{1}{2}\int_0^z \dif u \Big[\iu \sigma J_\mu(z-u)J_{-\mu -\alpha}(u) - J_\mu(z-u) J_{1-\mu-\alpha}(u) \Big] \nonumber\\ \approx& \frac{1}{\pi} \int_0^z \frac{\dif u}{\sqrt{u(z-u)}}\bigg[ \iu \sigma \cos\left(z - u - \frac{\mu \pi}{2} - \frac{\pi}{4}\right)\cos\left(u + \frac{(\mu+\alpha) \pi}{2} - \frac{\pi}{4}\right) -
			\cos\left(z - u - \frac{\mu \pi}{2} - \frac{\pi}{4}\right)\cos\left(u + \frac{(\mu+\alpha-1) \pi}{2} - \frac{\pi}{4}\right)\bigg] \nonumber\\
			&\approx \frac{1}{2\pi} \left[\iu \sigma \cos\left(z -\frac{\pi(1-\alpha)}{2} \right) - \cos\left(z - \frac{\pi(2 - \alpha)}{2}\right)  \right]\int_0^z \frac{\dif u}{\sqrt{u(z-u)}}\nonumber\\
			&= \frac{1}{2}e^{\iu \sigma( z + \pi\alpha/2)}
		\end{align}
	\end{widetext}
	where terms that integrate quickly with $u$ have been dropped. We thus obtain (recalling that $\sum_k \alpha_k$ is an integer and $\sum_k \gamma_k = 1$)
	\begin{align}
		\prod_k A^{(\alpha_k)}_k \approx e^{2\iu \sigma_i \sigma_f u_i u_f} (\sigma_i \sigma_f)^{\sum_k \alpha_k} \prod_k \cos\left(\pi \alpha_k \frac{1-\sigma_i\sigma_f}{2}\right)
	\end{align}
	The above is now manifestly invariant under shifts of $\alpha_k \rightarrow \alpha_k + n$, with $n \in \mathbb{Z}$, as we would expect. Note that when $\sigma_i = \sigma_f$, $A_k^{(\alpha_k)}$ becomes completely independent of $\alpha_k$, and so the difference of products in \eqref{eq:Lab} will vanish, leaving only the $\sigma_i = -\sigma_f$ terms, whereupon we can set $\prod_k A_k^{(\alpha_k)} = (-1)^{\sum_k \alpha_k} e^{-2\iu u_i u_f} \prod_k \cos(\pi \alpha_k)$. We now have
	\begin{align}
		I(v,t_2) &\approx -\left(\frac{\iu}{2\pi t_2}\right)^{N-2} \frac{2e^{-\iu Mv^2t_2/2}\prod_k m_k}{\pi \beta M^2} \Upsilon[\{\alpha_k\}] \nonumber\\ &\times\int_0^\infty u_i^{1/2} \dif u_i \int_0^\infty u_f^{1/2} \dif u_f e^{-\iu(u_i^2 + u_f^2)} \nonumber\\ &\sum_{\sigma_i = \pm 1} e^{\iu \pi \sigma_i/2}e^{-\iu\sigma_i \beta(x_f+x_i)}e^{-2\iu u_i u_f}
	\end{align}
	where the topological quantity $\Upsilon[\{\alpha_k\}]$ is given in Eq.~\eqref{eq:UpsilonDef}

	Now we make the transformation to variables $X = u_i + u_f$ and $x = u_f - u_i$, giving
	\begin{widetext}
		\begin{align}
			=& -\left(\frac{\iu}{2\pi t_2}\right)^{N-2} \frac{e^{-\iu Mv^2t_2/2}\prod_k m_k}{2\pi \beta M^2} \Upsilon[\{\alpha_k\}]\sum_{\sigma = \pm 1} (\iu \sigma) \int_0^\infty \dif X \int_{-X}^X \dif x \sqrt{X^2 - x^2} e^{-\iu X^2 - \iu \beta \sigma X} \nonumber\\
			=& -\iu \left(\frac{\iu}{2\pi t_2}\right)^{N-2} \frac{ e^{-\iu Mv^2t_2/2}\prod_k m_k}{4\beta M^2} \Upsilon[\{\alpha_k\}] \int_0^\infty \dif X\, X^2 e^{-\iu X^2} [e^{-\iu \beta X} - e^{\iu \beta X}]
			\label{eq:IPreInt}
		\end{align}
	\end{widetext}
	
	The above can be evaluated by defining
	\begin{align}
		&J(a,\beta) \coloneqq \int_0^\infty \dif X e^{-\iu a X^2} e^{-\iu \beta X} \nonumber\\ &= \frac{e^{\iu \beta^2/4a}}{\sqrt{a}}\Big[C(\infty) - C(\beta/2\sqrt{a}) - \iu S(\infty) + \iu S(\beta/2\sqrt{a})\Big]
	\end{align}
	where $C(x)$, $S(x)$ are the Fresnel integrals. This allows us to express the integral in question as $\iu\times\partial/\partial a [J(a,\beta) - J(a,-\beta)]|_{a=1}$, which evaluates to
	\begin{align}
		&\frac{-e^{\iu \beta^2/4}}{2} \Big[ (\beta^2 - 2\iu)\big(C(\beta/2) - \iu S(\beta/2)\big) - \iu \beta e^{-\iu \beta^2/4}\Big] \nonumber\\ &\xrightarrow{\beta \rightarrow \infty} -e^{\iu \beta^2/4}\frac{e^{-\iu\pi/4}\sqrt{\pi}\beta^2}{4}
	\end{align}
	Substituting the above into Eq.~\eqref{eq:IPreInt}, and restoring the original dimensionful quantities, we finally obtain Eq.~\eqref{eq:IvtFinal}.

	\section{Scaling of the average distance for particles with mutual anyonic statistics \label{app:AreaScaling}}
	
	In this appendix, following on from the discussion of Section \ref{subsec:MultipletBraiding}, we provide a more detailed proof that the area functional $A_c$ appearing Eq.~\eqref{eq:SumAreas} scales as $t_2^{3/2}$ even when there are non-trivial braiding phases between anyons created by the probe pulse. This confirms that the pump-probe response coefficient $\chi^{(3)}_{\rm PP}$, which we calculated explicitly in the absence of intra-multiplet interactions, continues to follow the universal relationship \eqref{eq:main-result} in this more general case.

	First, independently of $\alpha$, using the same geometric arguments as in Section \ref{subsec:PathInteg} we always expect a scaling relation of the form
	\begin{align}
		A_c \sim |\vec{v}| \int_0^{t_2} \dif \tau \Big\langle|{r}_{2, \perp}(\tau) - {r}_{1, \perp}(\tau)|\Big\rangle
		\label{eq:AcOriginal}
	\end{align}
	with $\vec{v}$ denoting the velocity of the pump anyon and $r_{j,\perp}$ ($j=1,2$) the component of the position of the $j$-th anyon along the direction perpendicular to $\vec{v}$. Here the angled brackets are a shorthand for the average over all paths $\vec{r}_{j}(t)$ that contribute to the two-time correlation function, i.e.~$\braket{C} \coloneqq (\int \pdif r_j(t) e^{\iu S} C)/(\int \pdif r_j(t) e^{\iu S})$ for any functional $C$. (Note that this is not necessarily a real quantity, but we are interested in the typical magnitude of $A_c$, and will therefore take an absolute value at the end.) The relation above was argued for in Subsec.~\ref{subsec:PathInteg} only based on the ballistic trajectory of the pump anyon, and the argument carries over to the case where anyons within a multiple have non-trivial mutual statistics.
	
	To compute the above, we will adopt the model of Ref.~\cite{Morampudi2017}, where the local operator $\mathcal{A}(\vec{r})$ creates the two anyons at a microscopic distance $a$ from one another. Specifically, $\ket{\Psi_{\mathcal{N}'}} = \int \dif^2 \vec{R} \int \dif \phi \ket{\vec{R}, a, \phi}_{\mathcal{N}'}$ where $\ket{\vec{R}, a, \phi}_{\mathcal{N}'}$ is a two-particle state with centre-of-mass coordinate $\vec{R}$ and relative separation $(a, \phi)$ in polar coordinates. Here $a$ is a UV length scale, which is required to regularize the state of the quasiparticles created by the operators $\hat{A}_{1,2}$ without violating the exclusion principle [previously given by Eq.~\eqref{eq:MultipletCreation}]. By dimensional analysis, the final result will be proportional to $t_2^{3/2}$ multiplied by a function of the dimensionless ratio $a\sqrt{m/t_2}$. The behaviour of this function at small arguments will determine the late-time scaling behaviour of $A_c$; the following calculation will demonstrate that this function tends to a constant as $a \rightarrow 0$, i.e.~the lengthscale $a$ falls out of the problem at late enough times.
	
	By translation invariance, the centre-of-mass coordinate and the relative coordinate decouple, and the statistical phases depend only on the latter. In fact, the part of the Hamiltonian controlling the motion of the relative coordinate is precisely the same as the transformed Hamiltonian $\hat{H}_k'$ appearing in Section \ref{subsec:Eval}, which describes a single particle orbiting around a flux tube of strength $2\pi \alpha$ at the origin. The eigenstates of this Hamiltonian are given in Eq.~\eqref{eq:Eigenstates}, and since the initial state $\ket{\Psi_{\mathcal{N}'}}$ is rotationally invariant we need only consider the zero angular momentum sector, $\ell = 0$. We have
	\begin{align}
		&\chi^{(1)}(t_2) \langle|{r}_{2, \perp}(\tau) - {r}_{1, \perp}(\tau)|\rangle = \int_0^{\infty} (\tilde{r} \dif \tilde{r}) \int_{-\pi}^\pi \dif \phi |\tilde{r}\cos \phi|  \nonumber\\
		&\times \braket{a, l=0|e^{-i(t_2-\tau)H}|\tilde{r},l=0} \braket{\tilde{r}, l=0|e^{-i\tau H}|a,l=0}
		\label{eq:AcExplicit}
	\end{align}
	To compute the necessary matrix elements, we require the expression for the eigenstates of $\hat{H}$, Eq.~\eqref{eq:Eigenstates}, along with the standard integral given in Ref.~\cite{DLMF}
	\begin{align}
		&\int_0^\infty x\dif x e^{-\iu p x^2} J_\nu(bx) J_\nu(cx) \nonumber\\ =& \frac{-\iu}{2p} e^{-\iu(b^2+c^2)/4p} e^{-\iu \pi \nu/2}J_\nu\left(\frac{bc}{2p}\right). & \Im p < 0
		\label{eq:BesselIntegral}
	\end{align}
	By setting $p = \tau/2m - \iu 0^+$ in the above, we find (leaving the infinitesimal imaginary shift implicit for convenience)
	\begin{align}
		&\braket{\tilde{r}, l=0|e^{-i\tau H}|a,l=0} \nonumber\\ =& \frac{-\iu m}{\tau} e^{\iu m (\tilde{r}^2+a^2)/2\tau} e^{-\iu \pi \alpha/2}J_{\alpha}\left(\frac{ma\tilde{r}}{\tau}\right).
		\label{eq:PropagatorRadial}
	\end{align}
	This can be substituted into \eqref{eq:AcExplicit}, and after evaluating the integral over the polar angle $\int_{-\pi}^\pi \dif \phi |\cos \phi| = 4$, we find
	\begin{align}
		\braket{A_c} &= \frac{1}{\chi^{(1)}(t_2)} \int_0^{t_2} \dif \tau \frac{-4m^2 e^{-\iu \alpha \pi}}{\tau(t_2 - \tau)}\int_0^\infty \tilde{r}^2 \dif \tilde{r} \nonumber\\ &\times e^{\iu m(\tilde{r}^2+a^2)(\tau^{-1} + (t_2-\tau)^{-1})/2} J_{\alpha}\left(\frac{ma\tilde{r}}{\tau}\right)J_{\alpha}\left(\frac{ma\tilde{r}}{t_2-\tau}\right) 
	\end{align}
	Notice that the integral over $\tau$ is unchanged upon making the transformation $\tau \rightarrow t_2 - \tau$; we can thus change the upper limit to $t_2/2$, and multiply the expression by 2. Defining the dimensionless parameter $\gamma \coloneqq ma^2/t_2$, we now transform to new integration variables $u = t_2/\tau$, $s = \gamma \tilde{r}/a$, giving
	\begin{align}
		\braket{A_c}\chi^{(1)}(t_2) &= -8\sqrt{mt_2}\gamma^{-3/2}\int_1^2 \dif u \frac{e^{-\iu \alpha \pi}}{u-1} \int_0^\infty s^2 \dif s\nonumber\\ &\times e^{\iu \gamma^{-1}(\gamma^2+s^2)u^2/(u-1)} J_\alpha\left( s \frac{u}{u-1}\right)J_\alpha (s u)
	\end{align}
	We are interested in the behaviour of this expression in the limit of small $\gamma \ll 1$. In this limit, the integrand is a fast-oscillating function of $s$, and so the integral will be dominated by contributions where $s \lesssim \gamma^{1/2}$. Since $u$ lies in the interval $[1,2]$, we can safely expand the second Bessel function for small arguments $J_\alpha(s u) \approx (s u/2)^\alpha/\Gamma(\alpha+1)$. (Note that the argument of the first Bessel function is large for $u$ close to 1, and therefore should not be expanded.) The integral over $s$ can be evaluated using another standard result \cite{Prudnikov1986}
	\begin{align}
		&\int_0^\infty \dif x\,x^{\mu-1}J_\alpha(b x) e^{-p x^2} \nonumber\\ =& \frac{(b/2)^\alpha}{2 p^{(\alpha + \mu)/2}} \frac{\Gamma((\alpha+\mu)/2)}{\Gamma(\alpha + 1)} {_1F_1}\left( \frac{\alpha+\mu}{2};\alpha + 1; \frac{-b^2}{4p}\right)
	\end{align}
	where ${_1F_1}(a;b;z)$ is the confluent hypergeometric function. Setting $\mu = 3 + \alpha$, $b = u/(u-1)$, and $p = -\iu \gamma^{-1} u^2/(u-1)$, we find
	\begin{align}
		&\braket{A_c}\chi^{(1)}(t_2) = -4\sqrt{mt_2}\gamma^{\alpha} \frac{e^{-\iu \alpha \pi/2+3\iu \pi/4}\Gamma(\alpha + 3/2)}{2^{2\alpha }\Gamma(\alpha+1)^2}\nonumber\\ \times& \int_1^2 \dif u \frac{\sqrt{u-1}}{u^3} e^{\iu \gamma u^2/(u-1)} {_1F_1}\left(\alpha + \frac{3}{2}; \alpha + 1; \frac{-\iu \gamma}{(u-1)}\right)
	\end{align}
	Since ${_1F_1}(a,b,-\iu x) = 1 + O(x)$ for small $x$, and $\sqrt{u-1}{_1F_1}(a+1/2,a,-\iu \gamma/(u-1))$ is bounded as $u \rightarrow 1$ for any $a$, the integral in the above converges to the constant $\pi/16$ as one takes the limit $\gamma \rightarrow 0$. Thus, we can read the time dependence off as $\braket{A_c}\chi^{(1)}(t_2) \propto t_2^{1/2 - \alpha}$. A simple calculation using Eq.~\eqref{eq:PropagatorRadial} gives the linear response coefficient as $\chi^{(1)}(t_2) \propto t_2^{-1-\alpha}$. Taking the ratio of these expressions, we see $\braket{A_c} \sim t_2^{3/2}$ as claimed.

	\bibliography{nonlinear_anyon}

\begin{thebibliography}{76}
\bibitem [{Per()}]{Perimeter}%
  \BibitemOpen
  \href@noop {} {}\bibinfo {note} {Note that, while formally the perimeter of
  typical Feynman path diverges in the continuum limit, here the correction
  scales like the area of a shell of width $r_{\rm int}$ surrounding the
  perimeter, which is finite.}\BibitemShut {Stop}%
\end{thebibliography}
	
\end{document}